\documentclass[reprint,
aps,
superscriptaddress
]{revtex4-2}

\usepackage{graphicx}
\usepackage{dcolumn}
\usepackage{bm}

\usepackage{amsmath,amssymb}
\usepackage{pifont}
\usepackage{float}
\usepackage{bm}
\usepackage[version=3]{mhchem}
\usepackage{comment}
\usepackage{cases}
\usepackage{framed}
\usepackage{wrapfig}
\usepackage{ascmac}
\usepackage{here}
\usepackage{array,booktabs}
\usepackage{color}
\usepackage{braket}
\usepackage{mathtools}
\usepackage[top=2cm, bottom=2cm, left=2cm, right=2cm]{geometry}

\usepackage{siunitx}

\graphicspath{{./image_main/}}

\newcommand{\n}{\nonumber \\}
\newcommand{\ceq}{\coloneqq}

\newcommand{\pd}{\partial}

\newcommand{\im}{\text{i}}
\newcommand{\nap}{\text{e}}

\begin{document}



\title{Fluxon Time-Delay Readout of a Superconducting Qubit Protected by a Spectral Gap in a Josephson Transmission Line}


\author{Shunsuke~Kamimura}
\email{shunsuke1997213@gmail.com}
\affiliation{Secure System Platform Research Laboratories, NEC Corporation, Kawasaki, Kanagawa, Japan}
\affiliation{NEC-AIST Quantum Technology Cooperative Research Laboratory, Tsukuba, Ibaraki 305-8568, Japan}

\author{Aree~Taguchi}
\affiliation{NEC-AIST Quantum Technology Cooperative Research Laboratory, Tsukuba, Ibaraki 305-8568, Japan}
\affiliation{Division of Physics, Faculty of Pure and Applied Sciences, University of Tsukuba, Tennodai, Tsukuba, Ibaraki 305-8571, Japan}

\author{Masamitsu~Tanaka}
\affiliation{Graduate School of Engineering, Nagoya University, Nagoya, Aichi 464-8603, Japan}

\author{Tsuyoshi~Yamamoto}%
 \email{tsuyoshi.yamamoto@aist.go.jp}
\affiliation{Secure System Platform Research Laboratories, NEC Corporation, Kawasaki, Kanagawa, Japan}
\affiliation{Division of Physics, Faculty of Pure and Applied Sciences, University of Tsukuba, Tennodai, Tsukuba, Ibaraki 305-8571, Japan}
\affiliation{Global Research and Development Center for Business by Quantum-AI technology (G-QuAT),
National Institute of Advanced Industrial Science and Technology (AIST), Tsukuba, Ibaraki, Japan}


\begin{abstract}

We theoretically investigate a readout scheme
of the quantum state of a superconducting qubit
based on time delay of a single flux quantum (SFQ),
also known as a fluxon, propagating in
a Josephson transmission line (JTL).
We concretely study the time-delay readout
based on capacitive coupling between
a transmon qubit and a JTL,
and we evaluate the time delay depending on the qubit state.
We also reveal a feature of the absence of fluxon pinning
and exponential suppression of nonadiabatic transitions
caused by the propagating fluxon,
which is advantageous for the time-delay readout.
We extend the analysis to a multi-level transmon as well.
Owing to the spectral gap in the JTL,
the radiative decay of the qubit mediated by the JTL
is exponentially suppressed,
and thus the transmission line itself
also serves as a filter protecting the qubit.
The readout scheme requires neither
complicated wiring to low-temperature stages
nor bulky microwave components,
which are bottlenecks for integration of
a large-scale superconducting quantum computer.

\end{abstract}

\maketitle


\section{\label{sec:Introduction}Introduction}

Superconducting quantum circuits are leading candidates
for a scalable quantum computer
since the first realization of a single
superconducting qubit~\cite{nakamura1999coherent},
and its various modifications have been
invented~\cite{mooij1999josephson,martinis2002rabi,koch2007charge,manucharyan2009fluxonium,leghtas2015confining}.
However, the hardware overhead is immense to demonstrate
computational advantage~\cite{shor1994algorithms,grover1997quantum,nielsen2010quantum}
of a large-scale fault tolerant quantum
computer~\cite{shor1995scheme,bravyi1998quantum,fowler2012surface,horsman2012surface},
due to required wiring to qubits and thermal-load overheads under
the brute-force scaling with current technology~\cite{mcdermott2018quantum}.
Especially for qubit readout,
which is one of the criteria for a scalable quantum computer~\cite{divincenzo2000physical},
today's systems largely rely on a near quantum-limited amplifier,
bulky microwave components such as circulators,
an amplifier based on high-electron-mobility transistor (HEMT),
and subsequent heterodyne detection
and thresholding at room temperature~\cite{gao2021practical},
which inevitably involves the significant overheads.

When reading out the quantum state of a superconducting qubit,
we often confront a trade-off that
strong coupling to an external apparatus
to accurately infer the quantum state
inevitably induces noise for the qubit.
For the most standard readout scheme referred to as
the dispersive readout~\cite{blais2004cavity,wallraff2004strong,schuster2007resolving},
a strong coupling between the qubit and a readout resonator
induces a large frequency shift of the resonator
depending on the qubit state,
while much strong coupling also gives rise to
fast radiative decay of qubit into an environment,
which significantly destroies
the qubit coherence~\cite{purcell1946spontaneous,fano1961effects,gardiner1985input,burkard2004multilevel,koshino2005quantum,blais2004cavity}.
To suppress the radiative decay into the environment,
one can utilize a filter,
refereed to as a Purcell filter~\cite{houck2008controlling,jeffrey2014fast,sete2015quantum},
that reduces the resonator density of states at the qubit frequency~\cite{blais2021circuit}.
In addition to such radiative decay,
unwanted state transitions
caused by microwave readout drive
inside or outside the qubit subspace,
or related state leakage referred to as qubit ionizations,
are being actively
investigated~\cite{thorbeck2024readout,bengtsson2024model,sank2016measurement,shillito2022dynamics,khezri2023measurement,dumas2024measurement,nesterov2024measurement,ye2024ultrafast,fechant2025offset,singh2025impact}.

Towards reailzation of a large-scale superconducting quantum computer,
one feasible approach is heterogeneous quantum-classical architecture
relying on the use of
superconducting
single-flux-quantum (SFQ)
based
digital logic,
such as rapid single flux quantum (RSFQ)~\cite{likharev1991rsfq},
energy-efficient RSFQ (ERSFQ)~\cite{kirichenko2011zero},
and ultra low-power, adiabatic quantum flux parametron (AQFP)~\cite{takeuchi2013adiabatic}
to mitigate the wiring and thermal-load
overheads~\cite{mcdermott2018quantum,mohseni2024build,bernhardt2025quantum,walter2025single,kawabata2026integration}.
On the one hand,
for a qubit control based on SFQ circuits,
theoretical proposals for single-qubit
and two-qubit gates are presented~\cite{mcdermott2014accurate,wang2023single,tonchev2025robust,torosov2025optimization,lapointe2025optimization},
and a single-qubit gate has already been experimentally demonstrated~\cite{leonard2019digital,liu2023single,liu2023quasiparticle}.
On the other hand,
in spite of theoretical~\cite{averin2006rapid,fedorov2007reading,herr2007design,naaman2013qubit,soloviev2015soliton,wustmann2026rapid}
and related experimental results~\cite{nakamura2009current,fedorov2014fluxon,van202228nm,ahmad2022scalable,das20244,lei2024cryogenic},
the experimental demonstration of a qubit readout scheme within
a low-temperature stage remains elusive.

We investigate a readout scheme based on time delay of a SFQ,
also called a fluxon,
caused by a superconducting qubit.
The time-delay readout scheme has been proposed~\cite{averin2006rapid}
and theoretically studied in detail~\cite{fedorov2007reading}
for inductive coupling between a flux qubit and a nonlinear
transmission line hosting a propagating fluxon,
which is called a Josephson transmission line (JTL).
We consider the time-delay readout of a transmon qubit~\cite{koch2007charge}
capacitively coupled to a JTL
and find analytical and numerical results of time delay
based on the perturbation theory of the JTL~\cite{keener1977solitons,mclaughlin1978perturbation}.
We reveal a distinct feature regarding the pinning effect of fluxon
for capacitive coupling to a transmon qubit;
a fluxon is not pinned by the capacitive coupling
even for an arbitrarily slow fluxon,
while there is a finite pinning velocity for the inductive coupling
below which the fluxon is pinned by
an effective potential~\cite{fedorov2007reading,aslamazov1984pinning}.
This feature of the capacitive coupling is advantageous
over the inductive coupling,
in that one can obtain a large amount of time delay
by slowering the initial velocity of the fluxon
without suffering from the pinning effect.
Also, as is done in the previous study~\cite{fedorov2007reading},
we estimate the degree of nonadiabatic transitions
caused by the capacitive coupling to the
JTL and find exponentially suppressed transition probabilities
with respect to the fluxon velocity.
We also consider a multi-level transmon and derive analytical results
of time delay depending on the photon-number states of the transmon,
and we discuss the absence of fluxon pinning
and unwanted state transitions inside and outside the qubit subspace.

Furthermore, we study radiative decay of a transmon qubit
mediated by a JTL.
Based on the observation that only small-amplitude oscillations are relevant
to typical low-impedance JTLs
possessing no fluxon~\cite{wildermuth2023quantum},
we approximate a JTL by a ``linearized'' but massive transmission line
that we call a Klein--Gordon transmission line (KGTL)
after the equation of motion it obeys.
By calculating the admittance of a dissipationless KGTL
seen by a capacitively coupled transmon,
we find an exponential suppression of the radiative decay rate
due to an external source.
This originates from a spectral gap appearing in the KGTL's
dispersion relation rooted in the mass term of the Klein--Gordon equation.
Thus, the JTL serves
not only as a medium transmitting the fluxon
but also as a filter protecting the qubit from the environment.
In typical circumstances,
even though we have a finite amount of the qubit decay rate
caused by a so-called subgap resistance
responsible for a leakage current of the JTL,
the amount of induced dissipation is relatively small
for typical underdamped JTLs,
and we can ignore its impact on the qubit coherence.

The reminder of this paper is organized as follows.
In Sec.~\ref{sec:model},
we provide classical Hamiltonians of the total system
composed of a transmon, a JTL, and a
coupling capacitor in between,
and the perturbation theory of JTL is briefly reviewed in this section.
Also, we address the quantum-mechanical treatment of the transmon
and typical values of circuit parameters.
In Sec.~\ref{sec:time_delay_qubit},
we give analytical and numerical results of the time delay
induced by capacitive coupling with a transmon qubit
under the two-level approximation.
In this section, we discuss the absence of pinning
and nonadiabatic unwanted state transitions,
and a method to enhance the readout precision via bias current
is theoretically and numerically examined.
We also presents typical time resolutions of delay detection
based on SFQ digital circuits.
In Sec.~\ref{sec:JTL_multilevel_transmon},
we extend the discussion to a multi-level transmon.
In Sec.~\ref{sec:radiative},
the degree of radiative decay is calculated for a dissipationless JTL
based on the approximation by a KGTL,
and also the decay rate for a dissipative but underdamped JTL
is evaluated to incorporate more practical circumstances.
Finally,
in Sec.~\ref{sec:discussion},
we provide discussions and perspectives of the present study,
and a conclusion follows in Sec.~\ref{sec:conclusion}.

\section{\label{sec:model}Model}

\subsection{\label{subsec:JTL_sineGordon}Josephson transmission line (JTL) and sine-Gordon equation}

\begin{figure}
	\begin{center}
		\includegraphics[clip,width=8cm,bb=0 0 575 225]{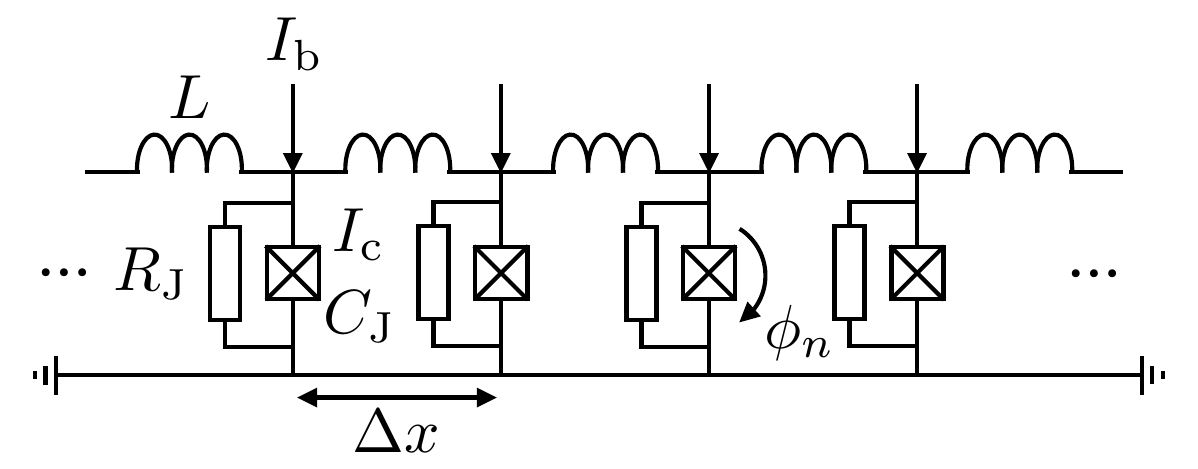}
		\caption{Schematic diagram of a discrete JTL, also called a lumped-element JTL,
		which is made of JJs connected each other via linear inductors.
		Each cell of length $\Delta x$ is characterized by the junction capacitance $C_{\text{J}}$,
		the junction critical current $I_{\text{c}}$,
		and the inductance $L$ of the upper electrode.
    Each junction is subject to a bias current $I_{\text{b}}$,
    and a resistor $R_{\text{J}} = 1/G_{\text{J}}$ describes normal current
    flowing across the junction.
		The discrete JTL
    gives a circuit model of a continuous JTL,
		which is also known as a long Josephson junction (LJJ).}
		\label{fig:JTL_R_phin}
	\end{center}
\end{figure}

A discrete JTL is made of Josephson junctions (JJs) connected each other
via linear inductors of an inductance $L$~(See Fig.~\ref{fig:JTL_R_phin}).
The JTL also has the capacitance $C_{\text{J}}$
and the Josephson energy
$E_{\text{J}}$ of JJs per unit cell of length $\Delta x$,
the latter of which is related to the junction critical current $I_{\text{c}}$,
as $E_{\text{J}} \ceq \varphi_0 I_{\text{c}}$.
Here,
$\varphi_0 \ceq \hbar / (2e) \simeq 3.29 \times 10^{-16}\,$Wb
denotes the reduced magnetic flux quantum,
which is defined by the reduced Planck constant~$\hbar = h / (2\pi)$
and the elementary charge~$e$.
From Kirchhoff's current and voltage laws,
we construct a classical Hamiltonian $H_{\text{JTL}}$ of the JTL, as
\begin{align}
	H_{\text{JTL}} &=
	\sum_{n=0}^N \left[
		\frac{Q_n^2}{2 C_{\text{J}}}
		+ \frac{\varphi_0^2}{2 L} (\phi_{n+1} - \phi_n)^2
		- E_{\text{J}} \cos \phi_n
	\right],
	\label{eq:discrete_JTL_Hamiltonian}
\end{align}
where the physical variables of the $n$-th junction are
the conjugate charge $Q_n$ to the flux $\Phi_n$ and
the dimensionless phase
$\phi_n \ceq \Phi_n / \varphi_0$.
The phase is positive when it increases in the path
across the junction from the upper electrode to the ground.
The time derivative of the phase $\phi_n$ is associated with the voltage
$V_n = - \varphi_0 \pd \phi_n / \pd t$
measured from the ground,
where the minus sign indicates the direction of the phase.
Under the Hamiltonian in Eq.~(\ref{eq:discrete_JTL_Hamiltonian}),
we derive the following equation of motion of JTL
from Hamilton's equations:
\begin{align}
	\frac{ \varphi_0 C_{\text{J}}}{ I_{\text{c}} } \frac{\pd^2 \phi_n}{\pd t^2}
	- \frac{\varphi_0}{L I_{\text{c}}} (\phi_{n+1} - 2 \phi_n + \phi_{n-1})
	+ \sin \phi_n &= 0.
	\label{eq:sG_discrete_JTL1}
\end{align}

In the continuum limit,
we take the lattice spacing $\Delta x$ to be zero
and promote the phase $\phi_n$ to a field $\phi (x,t)$
with circuit parameters per unit length
$c_{\text{J}} \ceq C_{\text{J}} /\Delta x,
i_{\text{c}} \ceq I_{\text{c}} / \Delta x$
and $\ell \ceq L / \Delta x$ being constant.
Then,
the equation of motion in Eq.~(\ref{eq:sG_discrete_JTL1})
leads to the following sine-Gordon equation:
\begin{align}
  \pd_{\tau}^2 \phi (\xi, \tau)
  - \pd_{\xi}^2 \phi (\xi, \tau)
  + \sin \phi (\xi, \tau)
  &= 0.
  \label{eq:sG_continuous_JTL1}
\end{align}
Here, we redefine the field $\phi (\xi, \tau)$ as a function of
the dimensionless coordinate $\xi \ceq x / \lambda_{\text{J}}$
and time $\tau \ceq \omega_{\text{p}} t$
using the Josephson penetration depth
$\lambda_{\text{J}} \ceq \sqrt{\varphi_0 / (\ell i_{\text{c}})}$
and the plasma frequency
$\omega_{\text{p}} \ceq \sqrt{ i_{\text{c}} / (\varphi_0 c_{\text{J}})}$.
Despite the nonlinearity of the sine-Gordon equation
in Eq.~(\ref{eq:sG_continuous_JTL1}),
this is an integrable partial differential equation
called a soliton equation,
and we can find its exact solutions.
One of solutions called a kink solution reads
\begin{align}
  \phi (\xi, \tau)
  &= 4 \, \text{arctan}
  \left[
    \exp \left( \pm \frac{\xi - u_0 \tau - \xi_0}{\sqrt{1 - u_0^2}} \right)
  \right],
  \label{eq:free_kink_expression}
\end{align}
where the constant and dimensionless parameters $u_0$ and $\xi_0$
describe the velocity and the initial position of the kink, respectively.
The parameters $u_0$ and $\xi_0$ are solely determined by
the initial conditions of the kink.
The sign $\pm$ represents the polarity of the kink,
which depends on the boundary condition of the system.
Although we assume the plus sign for concrete calculations,
an extension to the minus sign is straightforward.
The dimensionless velocity $u_0$ satisfies the inequality
$-1 < u_0 < 1$, which means that the speed of the kink
cannot exceed the velocity
$\overline{c} \ceq \lambda_{\text{J}} \omega_{\text{p}} = 1/\sqrt{\ell c_{\text{J}}}$.
This the reason why $\bar{c}$,
the Swihart velocity,
is also called the {\it speed of light} of the JTL.

For the kink solution of the plus polarity,
we can calculate the voltage
$V(\xi, \tau) = - \varphi_0 \omega_{\text{p}} \pd \phi / \pd \tau$
measured from the ground, as
\begin{align}
  V (\xi, \tau)
  &= \frac{\hbar \omega_{\text{p}}}{e} \,
  \frac{u_0}{\sqrt{1 - u_0^2}} \,
  \text{sech} \left(\frac{\xi - u_0 \tau - \xi_0}{\sqrt{1 - u_0^2}} \right),
\end{align}
where the hyperbolic function sech is defined as
$\text{sech} (x) \ceq 1/ ( \cosh (x)) = 2 /(\nap^x + \nap^{-x})$.
The voltage pulse is called a single-flux-quantum (SFQ) pulse.
Depending on the demensionless velocity $u_0$,
the height and the width of the voltage pulse vary,
respectively,
while the time integral is conserved:
\begin{align}
  \int_{-\infty}^{\infty} \! \text{d} t \, V
  &= 2\pi \varphi_0
	\simeq 2.07 \times 10^{-15} \, \text{Wb}.
\end{align}
This reflects the physical nature of the kink
as a quantized magnetic flux,
also called a fluxon.

We can apply a bias current $I_{\text{b}}$ to each junction
to modulate the velocity of the fluxon.
Also, in practical circumstances,
we have a finite amount of normal current across the junction,
which is modeled by a resistor $R_{\text{J}}$ shunting the junction
(See Fig.~\ref{fig:JTL_R_phin}).
We introduce the conductance $G_{\text{J}} = 1/R_{\text{J}}$
that quantifies the degree of normal current
of the JTL per unit cell,
which gives rise to dissipation.
Exploiting Kirchhoff's laws including those two circuit elements
and taking the continuum limit with
$g_{\text{J}} = G_{\text{J}}/ \Delta x$
and $i_{\text{b}} = I_{\text{b}} / \Delta x$ being constant,
we obtain the following perturbed sine-Gordon equation
with bias and dissipation terms:
\begin{align}
  \pd_{\tau}^2 \phi
  - \pd_{\xi}^2 \phi
  + \sin \phi
  &=
  \gamma
  - \alpha \pd_{\tau} \phi.
  \label{eq:sGeq_bias_dissipation}
\end{align}
The dimensionless parameters
are respectively defined as
$\alpha \ceq g_{\text{J}} \sqrt{ \varphi_0 / (i_{\text{c}} c_{\text{J}})}$
and $\gamma \ceq i_{\text{b}} / i_{\text{c}}$,
the former of which is called a damping coefficient.
Although the surface loss of the upper electrode of the JTL
can be modeled by an additional term $\beta \pd_{\tau \xi \xi} \phi$
to the right-hand side in Eq.~(\ref{eq:sGeq_bias_dissipation}),
we assume that its contribution
negligibly small in this study.
One can add bias current to
the JTL spatially nonuniformly,
and then the parameter $\gamma$ is no longer a constant
but a function $\gamma (\xi)$ of the coordinate $\xi$,
while we assume that it is constant otherwise mentioned.

One straightforward approach to solutions of the perturbed sine-Gordon equation
is discretizing the space by a mesh of spacing $a$
and numerically solving the following equations:
\begin{align}
  \pd_{\tau}^2 \phi_i
  - \frac{\lambda_{\text{J}}^2}{a^2} (\phi_{i+1} - 2 \phi_i + \phi_{i-1})
  + \sin \phi_i
  &= \gamma_i - \alpha_i \pd_{\tau} \phi_i,
  \label{eq:sG_eq_bias_dissipation_descritized}
\end{align}
where $\phi_i$ is the phase at each discretized point,
and $\gamma_i$ and $\alpha_i$ can depend on the points,
respectively.
Although this method can incorporate the nonuniformities
of the bias and the dissipation into the model,
the numerical simulations are expensive
and the results are not so clear to understand
in a simple way.

\subsection{\label{subsec:Perturbation}Perturbation theory of sine-Gordon equations}

Another possible approach is
resorting to the perturbation theory of sine-Gordon
equations~\cite{keener1977solitons,mclaughlin1978perturbation},
which is applicable to the situation
where the bias and dissipation terms are small,
i.e.~$|\gamma|, \alpha \ll 1$,
and we can treat them as perturbations.
Then, we assume the perturbative kink solution of the phase field
\begin{align}
  \phi (\xi, \tau) &=
  4 \, \text{arctan}
  \left[
    \exp
    \left(
      \frac{ \xi - \Xi (\tau) }{ \sqrt{1 - u^2 (\tau)}}
    \right)
  \right].
  \label{eq:modulated_solution_kink}
\end{align}
The time-dependent functions $u (\tau)$ and $\Xi (\tau)$
describe the velocity and the centroid of the kink, respectively,
and they are called modulation parameters.
According to the perturbation theory of the sine-Gordon soliton,
the bias and dissipation terms affect
the dynamics of the modulation parameters,
which are governed by the following ordinary differential equations:
\begin{align}
  \frac{\text{d} u}{\text{d} \tau}
  &= - \frac{1}{4} \pi \gamma (1-u^2)^{\frac{3}{2}}
  - \alpha u (1-u^2), 
  \label{eq:ode_modulation_u_sG} \\
  \frac{\text{d} \Xi}{\text{d} \tau}
  &= u.
  \label{eq:ode_modulation_Xi_sG}
\end{align}
Numerically solving the above differential equations
is much more feasible compared with the aforementioned
approach via direct discretization.
We can derive the steady-state velocity
$u_{\text{ss}} \ceq u (\infty)$
from the condition
$\text{d} u / \text{d} \tau = 0$ in Eq.~(\ref{eq:ode_modulation_u_sG}):
\begin{align}
  u_{\text{ss}} &=
  - \text{sign} (\gamma)
  \left[
    1 + \left(\frac{4 \alpha}{\pi \gamma}\right)^2
  \right]^{- \frac{1}{2}} ,
  \label{eq:uss_fluxon_velocity}
\end{align}
where $\text{sign} (\gamma)$ takes $1$ for $\gamma > 0$
and $-1$ for $\gamma < 0$.
We observe that the balance between the bias and the
dissipation determines the final velocity of the fluxon.

\subsection{\label{subsec:transmon}Transmon}

\begin{figure}
	\begin{center}
		\includegraphics[clip,width=8cm,bb=0 0 550 350]{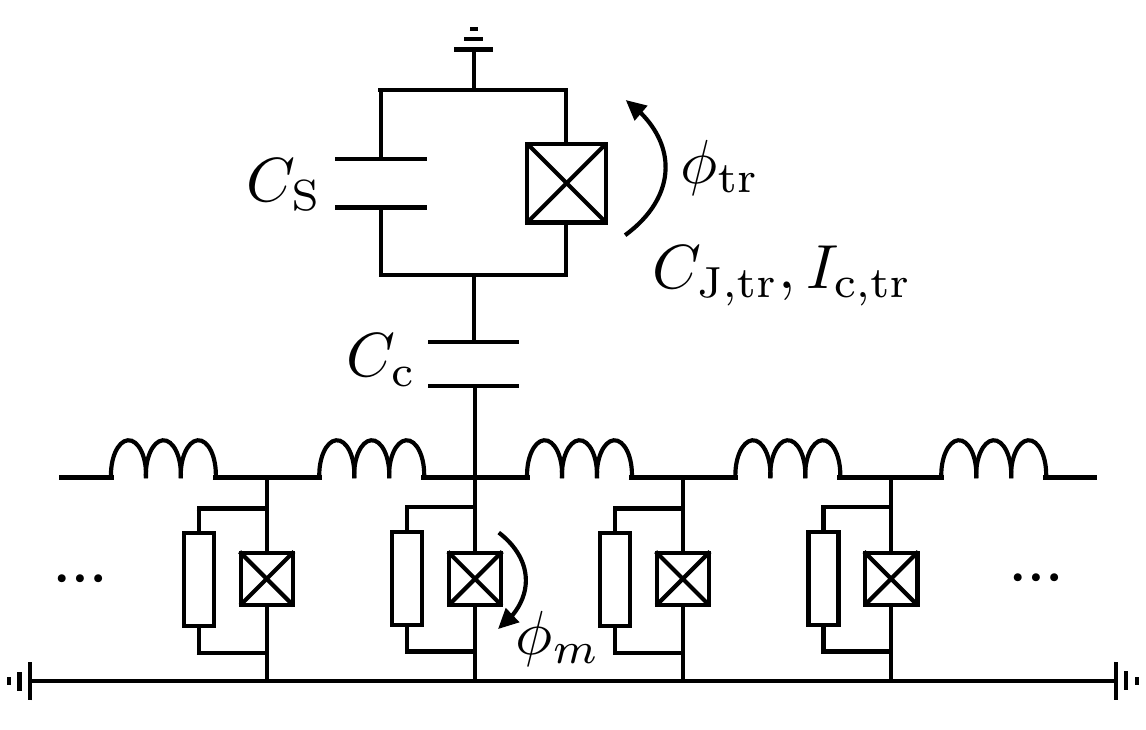}
		\caption{Schematic diagram of a transmon capacitively coupled to a JTL.
		The transmon is characterized by the total capacitance
		$C_{\Sigma} \ceq C_{\text{S}} + C_{\text{J,tr}}$
    ---the summation of the shunt capacitance $C_{\text{S}}$ and
		the transmon junction capacitance $C_{\text{J,tr}}$
    ---and the
    critical current $I_{\text{c,tr}}$
    of the junction of the transmon.
		The transmon charging and Josephson energies
		$E_{\text{C,tr}} \ceq e^2 / (2 C_{\Sigma}) $ and
		$E_{\text{J,tr}} \ceq \varphi_0 I_{\text{c,tr}}$
		determine the properties of the transmon.
		The transmon is coupled to the $m$-th junction of the JTL
		via the coupling capacitor $C_{\text{c}}$.}
		\label{fig:JTL_transmon_ccoupling}
	\end{center}
\end{figure}

A transmon qubit~\cite{koch2007charge}
is typically made of a JJ shunted by a large capacitor
that renders the qubit robust
against charge noise from an environment
(See Fig.~\ref{fig:JTL_transmon_ccoupling}).
According to Kirchhoff's laws and
the conventional quantization,
we obtain the quantum Hamiltonian $\hat{H}_{\text{tr}}$ of the transmon,
as
\begin{align}
  \hat{H}_{\text{tr}} &=
  4 E_{\text{C,tr}} \hat{n}_{\text{tr}}^2
  - E_{\text{J,tr}} \cos \hat{\phi}_{\text{tr}}.
\end{align}
The charging energy $E_{\text{C,tr}} \ceq e^2 / (2 C_{\Sigma})$
is defined by the transmon capacitance as the summation
$C_{\Sigma} \ceq C_{\text{J,tr}} + C_{\text{S}}$
of the shunt capacitance $C_{\text{S}}$
and the junction capacitance $C_{\text{J,tr}}$,
and $E_{\text{J,tr}} \ceq \varphi_0 I_{\text{c,tr}}$
denotes the Josephson energy of the transmon,
where $I_{\text{c,tr}}$ is the critical current.
The Cooper-pair number and phase operators
$\hat{n}_{\text{tr}} \ceq \hat{Q}_{\text{tr}} / (2e)$
and $\hat{\phi}_{\text{tr}} \ceq \hat{\Phi}_{\text{tr}} / \varphi_0$
denote the variables of the transmon, respectively,
which satisfy the commutation relation
$[\hat{\phi}_{\text{tr}}, \hat{n}_{\text{tr}}] = \im$.
After the conventional treatment~\cite{koch2007charge,blais2021circuit},
we obtain the easily diagonalizable form of Hamiltonian
under the condition
$E_{\text{J,tr}} / E_{\text{C,tr}} \simeq 50 \gg 1$,
as
\begin{align}
  \hat{H}_{\text{tr}}
  &= (\hbar \omega_{\text{tr}} - E_{\text{C,tr}}) \hat{a}^{\dagger} \hat{a}
  - \frac{E_{\text{C,tr}}}{2} \hat{a}^{\dagger} \hat{a}^{\dagger} \hat{a} \hat{a},
  \label{eq:transmon_Hamiltonian_nonlinearity}
\end{align}
where the bare transmon angular frequency reads
$\omega_{\text{tr}} \ceq \sqrt{ 8 E_{\text{J,tr}} E_{\text{C,tr}}} / \hbar$.
We introduced bosonic annihilation and creation operators
$\hat{a}$ and $\hat{a}^{\dagger}$
satisfying the commutation relation
$[\hat{a}, \hat{a}^{\dagger}] = 1$:
\begin{align}
  \hat{a} &\ceq \frac{1}{\sqrt{2}}
  \left(
    \frac{\hat{\phi}_{\text{tr}}}{\phi_{0,\text{tr}}}
    + \im \frac{\hat{n}_{\text{tr}}}{n_{0,\text{tr}}}
  \right).
\end{align}
The zero-point fluctuations of the phase and Cooper-pair number
$\phi_{0,\text{tr}} \ceq (8 E_{\text{C,tr}} / E_{\text{J,tr}})^{1/4}$
and
$n_{0,\text{tr}} \ceq \{ E_{\text{J,tr}} / (8 E_{\text{C,tr}}) \}^{1/4}$
are respectively defined.
These expressions give us
$\sqrt{2} \hat{\phi}_{\text{tr}} = \phi_{0,\text{tr}} (\hat{a} + \hat{a}^{\dagger})$
and
$ \sqrt{2} \hat{n}_{\text{tr}} = - \im n_{0,\text{tr}} (\hat{a} - \hat{a}^{\dagger})$.
We can diagonalize the transmon Hamiltonian
in Eq.~(\ref{eq:transmon_Hamiltonian_nonlinearity})
by the Fock state $\ket{n}$ such that
$\hat{a}^{\dagger} \hat{a} \ket{n} = n \ket{n}$
for non-negative integers $n$.
Explicitly,
\begin{align}
  \hat{H}_{\text{tr}}
  &= \sum_{n = 0}^{\infty} E_{n} \ket{n} \! \bra{n}, \\
  E_n &\ceq \hbar \omega_{\text{tr}} n
  - \frac{E_{\text{C,tr}}}{2} n (n+1).
  \label{eq:Energy_n_Fock_states}
\end{align}

For the two-level approximation,
we introduce the qubit states
$\ket{\uparrow} \ceq \ket{1}$
and
$\ket{\downarrow} \ceq \ket{0}$
of the transmon
and truncate the higher excited states
$\ket{2}, \ket{3}, \ket{4}, \ldots$
and so forth.
Then, the Hamiltonian of the transmon
is rewritten
with respect to the Pauli operator acting only onto the qubit subspace, as
\begin{align}
  \hat{H}_{\text{tr}} &=
  \frac{\hbar \omega_{\text{q}} }{2} \hat{\sigma}_z,
\end{align}
and we also have
$\sqrt{2} \hat{\phi}_{\text{tr}} = \phi_{0,\text{tr}} \hat{\sigma}_x$
and
$\sqrt{2} \hat{n}_{\text{tr}} = - n_{0,\text{tr}} \hat{\sigma}_y$.
Here, we define
$\hat{\sigma}_x \ceq\ket{\uparrow} \! \bra{\downarrow} + \ket{\downarrow} \! \bra{\uparrow}$,
$\hat{\sigma}_y \ceq -\im \ket{\uparrow} \! \bra{\downarrow} + \im \ket{\downarrow} \! \bra{\uparrow}$,
and
$\hat{\sigma}_z \ceq \ket{\uparrow} \! \bra{\uparrow} - \ket{\downarrow} \! \bra{\downarrow}$,
and $\omega_{\text{q}} \ceq E_1 / \hbar$ denotes
the angular frequency of the transmon qubit
under the two-level approximation.
The Pauli matrices satisfy the commutation relations
$[ \hat{\sigma}_{\alpha}, \hat{\sigma}_{\beta}]
= 2 \im \sum_{\gamma} \epsilon_{\alpha \beta \gamma} \hat{\sigma}_\gamma$,
where $\epsilon_{\alpha \beta \gamma}$ denotes the Edighnton's epsilon.

\subsection{\label{subsec:JTL_transmon_coupling}JTL-transmon capacitive coupling}

We consider a transmon capacitively coupled to a JTL
via a coupling capacitor of capacitance $C_{\text{c}}$
connecting the transmon and the $m$-th junction of the JTL
(See Fig.~\ref{fig:JTL_transmon_ccoupling}).
Based on Kirchhoff's laws, we obtain
the classical Hamiltonian of the total system, as
\begin{align}
  H
  &= H_{\text{JTL}} + H_{\text{tr}} + H_{\text{int}}.
\end{align}
The newly introduced interaction Hamiltonian
$H_{\text{int}}$ explicitly reads
\begin{align}
  H_{\text{int}} &\ceq
  \frac{C_{\text{c}}}{C_{\text{J}} C_{\Sigma}} Q_m Q_{\text{tr}}.
\end{align}
We here promote the variables only for the transmon to the operators
$\hat{\phi}_{\text{tr}}$
and $\hat{n}_{\text{tr}}$.
We treat the JTL classical mechanically
due to the fact that for a typical low-impedance JTL
satisfying $Z_{\text{J}} \ceq \sqrt{\ell / c_{\text{J}}}
\ll R_{\text{Q}} \ceq h / (2e)^2 \simeq 6.45 \,$k$\Omega$,
which is the focus of the present paper,
quantum fluctuations of the JTL are negligibly small~\cite{kato1996macroscopic,kemp2002josephson,wildermuth2022fluxons}.
In terms of the transmon operators,
the quantum Hamiltonians acting onto the transmon degrees of freedom
are written as
\begin{align}
  &\hat{H}_{\text{tr}} + \hat{H}_{\text{int}} \n
  &= \sum_{n=0}^{\infty} E_n \ket{n} \! \bra{n}
  - \im \sqrt{2} e n_{0,\text{tr}} 
  \frac{C_{\text{c}}}{C_{\Sigma} C_{\text{J}}}
  Q_m (\hat{a} - \hat{a}^{\dagger}).
  \label{eq:Htr_Hint_quantum_multi}
\end{align}

\subsection{\label{subsec:circuit_parameters}Circuit parameters}

We assume the values of the circuit parameters of
the total system, as shown in
Table~\ref{tab:circuit_params}.
The values are typical for an long Josephson junction (LJJ)
with the process rule
of the critical current density
$j_{\text{c}} = 250 \,$A/cm${}^2$
and the capacitance per unit area
$C_{\text{area}} = 60 \,$fF/$\mu$m${}^2$.
The width $w$ and the thickness $b$ of the upper electrode
and the length $l$ of the LJJ
are assumed to be
$w = 0.5 \, \mu$m, $b = 0.3 \, \mu$m, and $l = 1.0 \,$mm,
respectively.
The inductance $\ell$ per unit length is estimated by
the formula of the geometric inductance
$\ell = (\mu_0/2\pi) [\ln (2l /r - 1)]$
for a circle wire of the radius $r = \sqrt{w b/\pi}$,
where $\mu_0 = 4 \pi \times 10^{-7} \,$H/m
denotes the permeability of free space.
The value of the damping
coefficient $\alpha$ depends on the quality of the junction,
the environmental temperature,
the critical current density $j_{\text{c}}$, and so forth.
Typical values of the damping coefficient of an underdamped LJJ is around
$\alpha \simeq 1.0 \times 10^{-3}$~\cite{matsuda1982observation,matsuda1983fluxon},
whereas a large subgap registance
corresponding to the coefficient $\alpha \simeq 1.3 \times 10^{-7}$ is reported
for a small Al/AlO${}_{\text{x}}$/Al junction of area $S = 23 \, \mu$m${}^2$
at millikelvin temperature~\cite{gubrud2002sub}.
We assume the value from
$\alpha = 1.0 \times 10^{-6}$ to $1.0 \times 10^{-3}$
for concrete calculations,
and we leave its experimental realization for future work.

\begin{table}
  \setlength{\tabcolsep}{3pt}
  \caption{Circuit parameters used in the simulation.}
  \label{tab:circuit_params}
  \begin{ruledtabular}
  \begin{tabular*}{\columnwidth}{@{\extracolsep{\fill}}lcr@{}}
    Parameter & Symbol & Value \\
    \hline
    Inductance per unit length & $\ell$ & $1.8 \times 10^{-6} \, \mathrm{H/m}$ \\
    Capacitance per unit length &$c_{\text{J}}$ & $3.0 \times 10^{-8} \, \mathrm{F/m}$ \\
    Critical current per unit length &$i_{\text{c}}$ & $1.25 \, \mathrm{A/m}$ \\
    Josephson penetration depth & $\lambda_{\text{J}}$ & $12 \, \mu \mathrm{m}$ \\
    Plasma frequency & $\omega_{\text{p}}$ & $360 \, \mathrm{GHz}$ \\
    \hline
    Capacitance of transmon & $C_{\Sigma}$ & $100 \, \mathrm{fF}$ \\
    Critical current of transmon & $I_{\text{c,tr}}$ & $20 \, \mathrm{nA}$ \\
    Josephson energy (devided by $h$) & $E_{\text{C,tr}} / h$ & $190 \, \mathrm{MHz}$ \\
    Charging energy (devided by $h$) & $E_{\text{J,tr}} / h$ & $9.9 \, \mathrm{GHz}$ \\
    Transmon qubit frequency & $\omega_{\text{q}} / (2 \pi)$ & $3.7 \, \mathrm{GHz}$ \\
    \hline
    Coupling capacitance & $C_{\text{c}}$ & $10 \, \mathrm{fF}$ \\
    Dimensionless coupling constant & $\eta_{\text{c}}$ & $2.9 \times 10^{-3}$
  \end{tabular*}
  \end{ruledtabular}
\end{table}

\section{\label{sec:time_delay_qubit}Time-delay readout of a transmon qubit}

\subsection{\label{subsec:time_delay_qubit}Perturbative effect on JTL induced by transmon qubit
under the two-level approximation}

Under the two-level approximation of the transmon,
the Hamiltonians acting onto the transmon qubit subspace
are expressed by the Pauli matrices, as
\begin{align}
  \hat{H}_{\text{tr}} + \hat{H}_{\text{int}}
  &= \frac{\hbar \omega_{\text{q}} }{2} \hat{\sigma}_z
  + \hbar g_{\text{c}} (t) \hat{\sigma}_y.
  \label{eq:transmon_qubit_gc_driving}
\end{align}
Here, we define the coupling amplitude
\begin{align}
  g_{\text{c}} (t)
  &\ceq
  - \sqrt{2} \, n_{0,\text{tr}} \frac{e}{\hbar}
  \frac{C_{\text{c}}}{C_{\Sigma} C_{\text{J}}}
  Q_m (t),
\end{align}
which is related to the voltage of the $m$-th junction
$V_{m} (t) = - Q_m (t) / C_{\text{J}}$
measured from the ground.
To derive an effective Hamiltonian
for a weak coupling
$|g_{\text{c}} (t)| \ll \omega_{\text{q}}$,
we exploit the Schrieffer--Wolff (SW) transformation.
Then, we obtain the following effective Hamiltonian
for a weak coupling with a slowly moving fluxon
of the initial velocity $u_0$
satisfying $u_0 \ll \omega_{\text{q}} / \omega_{\text{p}}$
(See Appendix~\ref{appsubsec:SW_qubit}):
\begin{align}
  \hat{H}_{\text{SW}} (t) &\simeq
  \frac{\hbar \omega_{\text{q}} }{2} \hat{\sigma}_z
  + \frac{ \hbar g_{\text{c}}^2 (t)}{\omega_{\text{q}}} \hat{\sigma}_z .
  \label{eq:SW_Hamiltonian_qubit}
\end{align}
The second term on the right-hand side of Eq.~(\ref{eq:SW_Hamiltonian_qubit})
is the effective perturbation term of the JTL.
This result corresponds to that derived
from the adiabatic approximation adopted
for inductive coupling between a JTL and a flux qubit
at the symmetry point~\cite{fedorov2007reading}.
Also, we note that the qubit free part
in the first term of Eq.~(\ref{eq:SW_Hamiltonian_qubit})
and the effective perturbation term commute,
which is the feature of
the non-demolition readout~\cite{blais2004cavity,blais2021circuit}.
The degree of nonadiabatic state transitions caused by the
fluxon is evaluated in subsection~\ref{subsec:qubit_nonadiabatic}.

Because the transmon parts are now written
solely by the Pauli operator $\hat{\sigma}_z$,
we treat $\hat{\sigma}_z$ as a c-number
$\sigma_z$ that takes $+1$ for $\ket{\uparrow}$
and $-1$ for $\ket{\downarrow}$.
Then, after the fluxon-qubit scattering,
the total Hamiltonian in the original frame reads
\begin{align}
  H &= H_{\text{JTL}} + H_{\text{SW}} (t) \n
  &= H_{\text{JTL}}
  + 2 n_{0,\text{tr}}^2 \frac{e^2}{\hbar \omega_{\text{q}}}
  \frac{C_{\text{c}}^2}{C_{\Sigma}^2 C_{\text{J}}^2} Q_m^2 \sigma_z
  + \frac{\hbar \omega_{\text{q}} }{2} \sigma_z .
\end{align}
Only retaining the Hamiltonians regarding the JTL,
we obtain the modified classical Hamiltonian $H_{\text{JTL}}'$
that describes a perturbed dynamics of JTL by the transmon qubit, as
\begin{align}
  H_{\text{JTL}}' 
  &\ceq
  H_{\text{JTL}}
  + 2 n_{0,\text{tr}}^2 \frac{e^2}{\hbar \omega_{\text{q}} }
  \frac{C_{\text{c}}^2}{C_{\Sigma}^2 C_{\text{J}}^2} Q_m^2 \sigma_z .
  \label{eq:H_JTL_prime_twolevel}
\end{align}
The perturbation term is nothing but a local change in the capacitance
of JTL's junction labeled by $m$,
and thus the resulting sine-Gordon equation involves an additional term
proportional to $\pd_{\tau}^2 \phi_m$.
In the continuum limit, we obtain
\begin{align}
  \pd_{\tau}^2 \phi
  + \alpha \pd_{\tau} \phi
  - \pd_{\xi}^2 \phi
  + \sin \phi
  &= \eta_{\text{c}} \delta (\xi) \sigma_z \pd_{\tau}^2 \phi,
  \label{eq:perturbed_sG_qubit}
\end{align}
where we set the origin of coordinate $\xi$ to be the position
of the $m$-th junction of the JTL
(See Appendix~\ref{appsubsec:perturbation_qubit}).
The dimensionless constant $\eta_{\text{c}} \ll 1$
of the capacitive coupling is defined as
\begin{align}
  \eta_{\text{c}} &\ceq
  \frac{1}{1 - p} \frac{C_{\text{c}}^2 }{ \lambda_{\text{J}} c_{\text{J}} C_{\Sigma}},
\end{align}
where $p$ is a dimensionless constant:
\begin{align}
	p &\ceq \sqrt{ \frac{ E_{\text{C,tr}} }{ 8 E_{\text{J,tr}}} }
  = \frac{1}{8} \phi_{0,\text{tr}}^2 .
\end{align}
For a typical transmon
with $E_{\text{J,tr}} / E_{\text{C,tr}} \simeq 50$,
$p \simeq 1/20$.

\subsection{\label{subsec:qubit_time_delay}Time delay of fluxons}

We analytically calculate the
time delay between the scattered fluxons
for $\ket{\uparrow}$ and $\ket{\downarrow}$.
Based on the perturbation theory of
the sine-Gordon kink~\cite{keener1977solitons,mclaughlin1978perturbation},
the velocity $u (\tau)$ and centroid $\Xi (\tau)$ in the perturbed kink
in Eq.~(\ref{eq:modulated_solution_kink})
obey the following ordinary differential equations
(See Appendix~\ref{appsubsec:Deqs_u_Xi}):
\begin{align}
  \frac{\text{d} u}{\text{d} \tau}
  &= - \alpha u (1 - u^2)
  - \frac{1}{2} \eta_{\text{c}} \sigma_z u^2
  \text{sech}^3 (\Theta_0) \sinh (\Theta_0),
  \label{eq:ODE_u_qubit} \\
  \frac{\text{d} \Xi}{\text{d} \tau}
  &= u + \frac{1}{2} \eta_{\text{c}} \sigma_z \frac{u^3}{\sqrt{1 - u^2}}
  \Theta_0 \, \text{sech}^3 (\Theta_0) \sinh (\Theta_0),
  \label{eq:ODE_Xi_qubit}
\end{align}
where we do not apply bias current to the JTL.
Here, we introduced the time-dependent function $\Theta_0$, as
\begin{align}
  \Theta_0 (\tau) &\ceq
  \frac{\Xi (\tau)}{\sqrt{1 - u^2 (\tau)}}.
\end{align}

We consider the situation in which a moving fluxon of the initial velocity
$u_0$ is scattered by the transmon qubit in the state
$\ket{\uparrow}$ or $\ket{\downarrow}$.
Then, the modulated velocity $u (\tau)$ and centroid $\Xi (\tau)$
are related with each other as
\begin{align}
  \sqrt{ \frac{1 - u^2}{1 - u_0^2} }
  &=
    1 - \frac{1}{4} \eta_{\text{c}} \sigma_z \frac{u_0^2}{\sqrt{1 - u_0^2}} \,
    \text{sech}^2 (\Theta_0),
  \label{eq:analytical_u_qubit}
\end{align}
where we assumed that the transmon-JTL interaction is more dominant
compared with the dissipation
(See Appendix~\ref{appsubsec:pert_sol_capa}).
Namely, we assume that
\begin{align}
  \alpha \ll \eta_{\text{c}} u_0.
  \label{eq:friction_small_cond}
\end{align}
For a fluxon propagating to the positive direction,
i.e. $u > 0$,
we can solve Eq.~(\ref{eq:analytical_u_qubit}) for $u$ and obtain
\begin{align}
  u (\Xi) &\simeq
  u_0 \left[
    1 + \frac{1}{4} \eta_{\text{c}} \sigma_z
    \sqrt{1 - u_0^2} \, \text{sech}^2
    \left(\frac{\Xi}{\sqrt{1 - u_0^2}}\right)
  \right],
\end{align}
where we neglected the higher-order terms.

Now, we are ready to analytically calculate
the time delay of the fluxon induced by the transmon qubit
under the two-level approximation.
Comparing the propagation times
$\tau_{\ket{\uparrow}}$ and $\tau_0$,
which respectively represent the propagation time under
the capacitive coupling
to the transmon qubit in the state $\ket{\uparrow}$
and under no coupling,
we calculate
\begin{align}
  \tau_{\ket{\uparrow}} - \tau_0
  &\ceq \int_{-\infty}^{\infty} \text{d} \Xi
  \left.\left(\frac{1}{u(\Xi)} - \frac{1}{u_0}\right) \right|_{\sigma_z = +1} \n
  &\simeq - \eta_{\text{c}} \frac{\sqrt{1 - u_0^2}}{4 u_0}
  \int_{-\infty}^{\infty} \! \text{d} \Xi \, \text{sech}^2
  \left(\frac{\Xi}{\sqrt{1 - u_0^2}}\right) \n
  &= - \eta_{\text{c}} \frac{1 - u_0^2}{2 u_0} .
\end{align}
We evaluated the integral by the method of residues.
The result means that the capacitive coupling
with the transmon qubit in the state
$\ket{\uparrow} = \ket{1}$ accelerates the fluxon.
Similarly,
\begin{align}
  \tau_{\ket{\downarrow}} - \tau_0
  &= \eta_{\text{c}} \frac{1 - u_0^2}{2 u_0},
\end{align}
which means that the transmon qubit in $\ket{\downarrow} = \ket{0}$
decelerates the fluxon.
Comparing the cases for $\ket{\uparrow}$ and $\ket{\downarrow}$,
the dimensionless time delay
$\tau_{\text{d}} \ceq \tau_{\ket{\downarrow}} - \tau_{\ket{\uparrow}}$
reads
\begin{align}
  \tau_{\text{d}} &\simeq
  \eta_{\text{c}} \frac{1 - u_0^2}{u_0}
  = \frac{1 - u_0^2}{u_0 (1-p)}
  \frac{C_{\text{c}}^2}{\lambda_{\text{J}} c_{\text{J}} C_{\Sigma}}.
  \label{eq:taud_qubit_ana01}
\end{align}
Recovering the dimension of time $\omega_{\text{p}}^{-1}$,
we end up with the time delay $T_{\text{d}}$ in a unit of time, as
\begin{align}
  T_{\text{d}}
  &\ceq
  \tau_{\text{d}} \omega_{\text{p}}^{-1}
  \simeq \frac{1 - u_0^2}{u_0 (1-p)}
  \frac{C_{\text{c}}^2}{C_{\Sigma}} Z_{\text{J}},
  \label{eq:Td_qubit_ana01}
\end{align}
where $Z_{\text{J}} = \sqrt{\ell / c_{\text{J}}}$
denotes the characteristic impedance of the JTL.
This expression illustrates one important consequence that
the fluxon intial velocity $u_0$ is a knob
that we can control to enhance the time delay
without changing the device parameters.
We also observe that we can enhance the time delay by utilizing a JTL
of a large characteristic impedance $Z_{\text{J}}$,
which can be increased
by exploiting the upper electrode made of
high-kinetic-inductance materials~\cite{wildermuth2022fluxons}.

\begin{figure}
	\begin{center}
		\includegraphics[clip,width=8cm,bb=0 0 550 340]{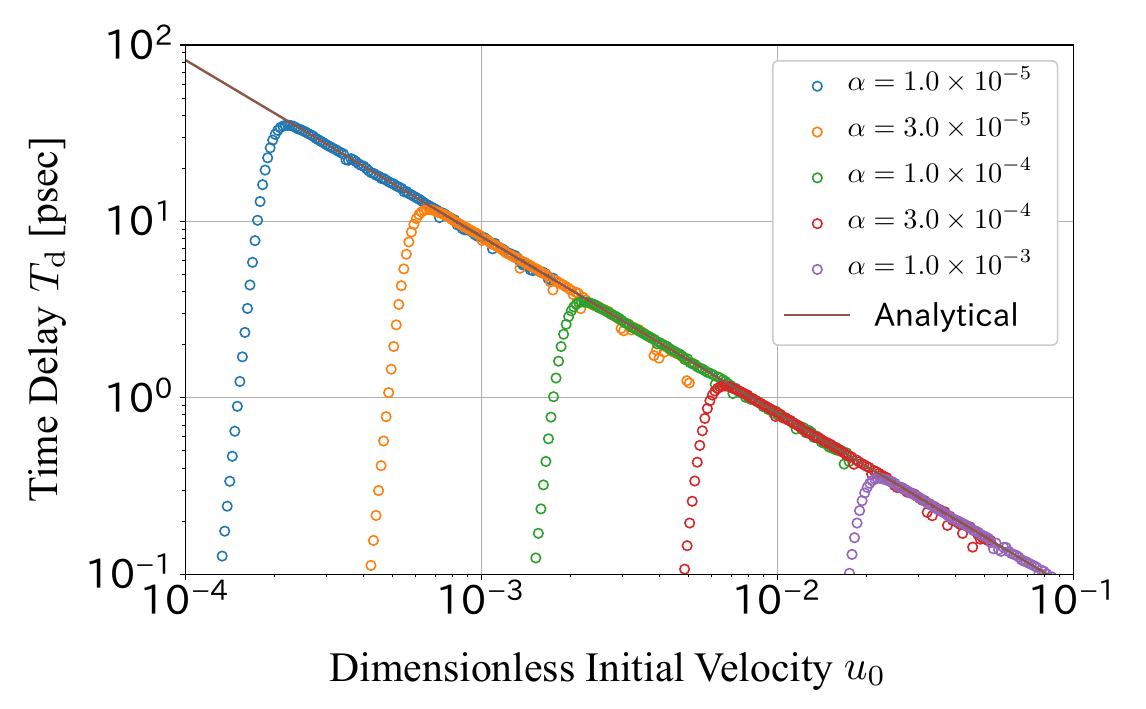}
		\caption{(circles) Numerical results of the time delay
    $T_{\text{d}} = (\tau_{\ket{\downarrow}} - \tau_{\ket{\uparrow}}) \omega_{\text{p}}^{-1}$
    calculated by numerically solving Eqs.~(\ref{eq:ODE_u_qubit}) and (\ref{eq:ODE_Xi_qubit})
    for several values of the damping coefficient $\alpha$
    and (line) analitycal result in Eq.~(\ref{eq:Td_qubit_ana01}).
    We adopt the values of circuit parameters in
    Table~\ref{tab:circuit_params}.}
		\label{fig:tr_qubit_time_delay}
	\end{center}
\end{figure}

In Fig.~\ref{fig:tr_qubit_time_delay},
we provide numerical results of the time delay
$T_{\text{d}}$,
obtained by numerically solving the differential equations of
the modulation parameters
in Eqs.~(\ref{eq:ODE_u_qubit}) and (\ref{eq:ODE_Xi_qubit}).
The values of parameters are presented in
Table~\ref{tab:circuit_params}.
For varied fluxon initial velocity $u_0$,
we confirm good agreements with the analytical results
for small values of damping coefficient $\alpha$.
However, as $\alpha$ increases, the numerical results deviate
from the analytical one below the threshold velocity,
which increases with $\alpha$.
This reflects the violation of the assumption
in Eq.~(\ref{eq:friction_small_cond}).
Therefore, a small damping coefficient is advantageous for
obtaining a large amount of time delay.

\subsection{\label{subsec:qubit_pinning}Absence of fluxon pinning}

As is done in the previous studies~\cite{aslamazov1984pinning,fedorov2007reading},
we examine the fluxon pinning effect induced by the transmon qubit
capacitively coupled to the JTL.
When the fluxon pinning occurs,
the velocity satisfies $u = 0$
at a certain position of centroid $\Xi$.
Supposing this condition in Eq.~(\ref{eq:analytical_u_qubit})
leads us to
\begin{align}
  \text{sech}^2 \Xi &= - \frac{2}{\eta_{\text{c}}} \sigma_z,
\end{align}
where we neglected the higher-order terms $O (\eta_{\text{c}}^2)$.
Therefore, under the condition of the perturbation theory
$\eta_{\text{c}} \ll 1$,
the pinning condition above
cannot be held for any $\Xi \in \mathbb{R}$
regardless of the value of $\sigma_z$.
We thus conclude that the fluxon pinning is absent
for weak capacitive coupling
between the transmon qubit and JTL,
which is distinct feature from that
for inductive coupling~\cite{aslamazov1984pinning,fedorov2007reading}.

\subsection{\label{subsec:qubit_nonadiabatic}Nonadiabatic transition}

We recall the Hamiltonian of the transmon qubit
under the two-level approximation
driven by the moving fluxon in Eq.~(\ref{eq:transmon_qubit_gc_driving}),
$\hat{H} (t) = \hat{H}_0 + \hat{V} (t)$,
where
$\hat{H}_0 \ceq (\hbar \omega_{\text{q}} / 2) \hat{\sigma}_z$
and
$\hat{V} (t) \ceq \hbar g_{\text{c}} (t) \hat{\sigma}_y$.
Because the driving term proportional to $\hat{\sigma}_y$
does not commute with the energy basis $\hat{\sigma}_z$ of the transmon,
we expect nonadiabatic transitions between
$\ket{\uparrow}$ and $\ket{\downarrow}$
induced by the driving.
The profile of the coupling amplitude $g_{\text{c}} (t)$
is explicitly given as
\begin{align}
  g_{\text{c}} (t) &=
  \sqrt{2} n_{0,\text{tr}} \frac{C_{\text{c}}}{C_{\Sigma}}
  \frac{u_0 \omega_{\text{p}}}{\sqrt{1 - u_0^2}} \,
  \text{sech} \left[
    \frac{u_0 \omega_{\text{p}} t + \xi_0}{\sqrt{1 - u_0^2}}
  \right].
  \label{eq:explicit_gc_profile}
\end{align}
Here,
the origin $\xi = 0$ denotes the position of the transmon qubit
and $\xi_0$ is defined as the
initial position of fluxon's centroid.

Based on Fermi's golden rule applicable for a weak coupling
$|g_{\text{c}} (t) | \ll \omega_{\text{q}}$,
we can calculate the probabilities
$P(\uparrow \! | \! \downarrow )$ and $P(\downarrow \! | \! \uparrow )$
of the transitions
$\ket{\downarrow} \mapsto \ket{\uparrow}$
and $\ket{\uparrow} \mapsto \ket{\downarrow}$
respectively, as
\begin{align}
  P(\uparrow \! | \! \downarrow ) &\simeq
  \frac{1}{\hbar^2}
  \left|
    \int_{-\infty}^{\infty} \! \text{d} s \,
    \nap^{ \im \omega_{\text{q}} s}
    \bra{\uparrow} \hat{V} (s) \ket{\downarrow}
  \right|^2, \\
  P(\downarrow \! | \! \uparrow ) &\simeq
  \frac{1}{\hbar^2}
  \left|
    \int_{-\infty}^{\infty} \! \text{d} s \,
    \nap^{ - \im \omega_{\text{q}} s}
    \bra{\downarrow} \hat{V} (s) \ket{\uparrow}
  \right|^2.
\end{align}
Utilizing the expression of $\hat{V} (s)$
and the formula
\begin{align}
  \int_{-\infty}^{\infty} \! \text{d} x \, \nap^{\im k x} \text{sech} (ax)
  &= \frac{\pi}{|a|} \text{sech} \left(\frac{\pi k}{2 a} \right),
  \label{eq:sech_Fourier_formula}
\end{align}
we obtain
\begin{align}
  P(\uparrow \! | \! \downarrow ) &=
  P(\downarrow \! | \! \uparrow ) \n
  &= 2 \pi^2 n_{0,\text{tr}}^2 \frac{C_{\text{c}}^2}{C_{\Sigma}^2}
  \text{sech}^2
  \left(\frac{\pi}{2} \frac{\sqrt{1 - u_0^2} \omega_{\text{q}}}
  {u_0 \omega_{\text{p}}}\right) .
  \label{eq:nonad_transition_prob_qubit}
\end{align}
From this result, the nonadiabatic transitions are exponentially suppressed
under the following condition on the fluxon
initial velocity $u_0$:
\begin{align}
  \frac{\sqrt{1 - u_0^2}}{u_0}
  &\gg \frac{2 \omega_{\text{p}}}{\pi \omega_{\text{q}}} .
\end{align}
In practical situations, the ratio is approximated as
\begin{align}
  \frac{2 \omega_{\text{p}}}{\pi \omega_{\text{q}} }
  &\simeq
  \frac{2 \times 360 \, \text{GHz}}{ \pi \times 2\pi \times 3.7 \, \text{GHz}}
  \simeq 9.9
\end{align}
and thus the nonadiabatic transitions are negligibly small
for the fluxon initial velocity $u_0 < 0.01$,
as is assumed to yield an enough amount of the time delay.
In our scheme employing the capacitive coupling,
we can decrease the initial velocity of fluxon
without suffering from the pinning caused by
the coupling to the qubit,
and thereby we can enhance the delay time
and suppress the nonadiabatic transitions at the same time.

\subsection{\label{subsec:pulse_separation}Pulse separation via bias current}

\begin{figure*}
\centering
		\includegraphics[clip,width=\textwidth,bb=0 0 700 550]{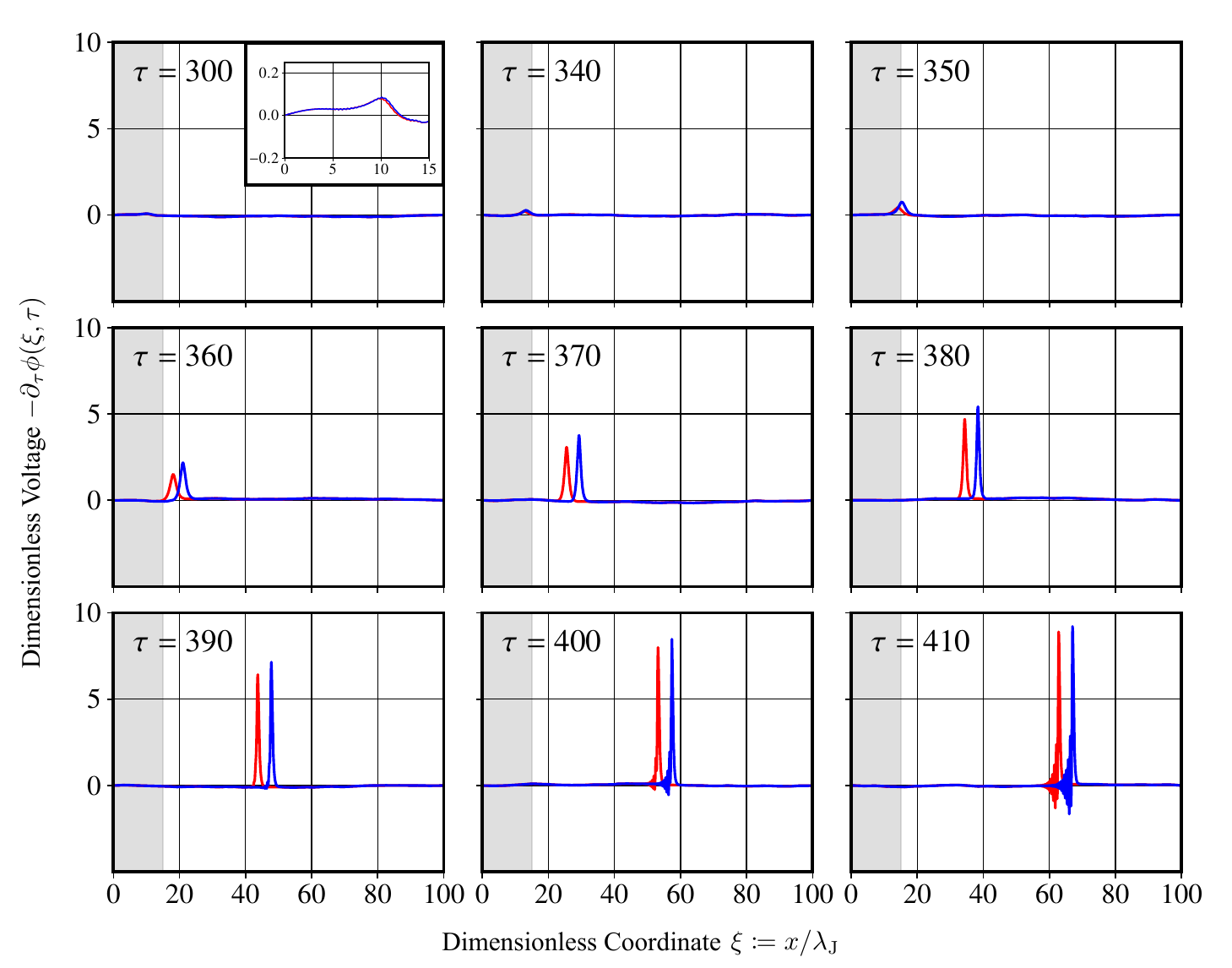}
		\caption{Numerically obtained snapshots of accelerated fluxons as
    dimensionless voltage pulses
    $V (\xi, \tau) / (\varphi_0 \omega_{\text{p}}) = - \pd_{\tau} \phi (\xi, \tau)$
    for two different initial positions of centroid
    $\xi_0 = 5.75$ and $5.85$.
    The grey area satisfying
    $0 \le \xi < 15$
    represents the nonbiased region,
    while the white area satisfying $\xi \ge 15$
    represents the biased region
    by the dimensionless bias $\gamma = -0.1$.
    The initial velocity is
    $u_0 = 0.01$ for both cases,
    and the damping coefficient is $\alpha = 1.0 \times 10^{-6}$.
    We assume the closed boundary conditions at the ends
    of the JTL,
    i.e. $\phi (0,\tau) = 0$ and
    $\phi (100, \tau) = 2\pi$ for an arbitrary time $\tau$.
    When we assume the values of circuit parameters in
    Table~\ref{tab:circuit_params},
    the unities of time $\tau$, coordinate $\xi$,
    and voltage are
    $2.8 \,$ps, $12 \, \mu$m, and $0.12 \,$mV,
    respectively.}
		\label{fig:pulse_separation_total}
\end{figure*}

Whereas we can increase the time delay $\tau_{\text{d}}$,
namely the temporal width of delay,
by modulating the fluxon initial velocity $u_0$,
the dimensionless spatial width of delay
$\Delta \xi_{\text{d}} \ceq u_0 \tau_{\text{d}}$
is constant under the change in $u_0$:
\begin{align}
  \Delta \xi_{\text{d}} &\simeq \eta_{\text{c}} \ll 1.
\end{align}
The unit of coordinate $\xi$ was $\lambda_{\text{J}}$
that is the spatial width of a slow fluxon,
thus the scattered fluxons for $\ket{\downarrow}$
and $\ket{\uparrow}$ are almost completely
overlapped each other,
which makes it hard to detect the delay between them
at the subsequent stage.
We could bump the fluxon $N_{\text{int}}$ times into the transmon
and obtain $N_{\text{int}}$ times larger spatial width of delay
$\Delta \xi_{\text{d,tot}} = N_{\text{int}} \eta_{\text{c}}$
by exploiting an annular LJJ~\cite{ustinov1992dynamics,le2025microwave}.

We here propose another approach based on bias current
after the fluxon scattering.
We concretely consider the following sine-Gordon equation
in which a nonuniform bias current $\gamma (\xi)$ is applied:
\begin{align}
  \pd_{\tau}^2 \phi
  - \pd_{\xi}^2 \phi
  + \sin \phi
  &=
  \gamma (\xi)
  - \alpha \pd_{\tau} \phi .
  \label{eq:sGeq_nonuni_bias}
\end{align}
In spite of the nonuniform bias term,
the equation of motion preserves the time-translation symmetry;
if a function $\phi = \phi (\xi, \tau)$ is a solution of
Eq.~(\ref{eq:sGeq_nonuni_bias}),
then another temporally delayed function
$\phi_{\text{d}} (\xi, \tau) = \phi (\xi, \tau - \tau_{\text{d}})$
is also a solution of Eq.~(\ref{eq:sGeq_nonuni_bias}).
This means that the time delay $\tau_{\text{d}}$ is preserved
before and after the nonuniform bias $\gamma (\xi)$.
On the other hand,
the space-translation symmetry is violated due to the nonuniform bias,
and thus the spatial width of delay is not preserved.
For simplicity, we consider the bias profile $\gamma (\xi)$ such that
$\gamma (\xi) = 0$ for $0 \le \xi < \xi_{\text{b}}$ and
$\gamma (\xi) = \gamma = \text{const.}$ for $\xi_{\text{b}} \le \xi$,
where $\xi_{\text{b}}$ denotes starting point of the biased region.
After the accelerated fluxon reaches the steady-state velocity
in Eq.~(\ref{eq:uss_fluxon_velocity}),
the spatial width of the fluxon becomes $\sqrt{1 - u_{\text{ss}}^2}$.
Comparing the spatial width $\sqrt{1 - u_{\text{ss}}^2}$
of the accelerated fluxon
and the amplified spatial width of delay $ u_{\text{ss}} \tau_{\text{d}}$,
we find the condition on the absolute value of the bias $|\gamma|$
for significant separation of the two fluxons
for $\ket{\downarrow}$ and $\ket{\uparrow}$:
\begin{align}
  |\gamma| & \gg \frac{ \alpha u_0}{\eta_{\text{c}}}.
\end{align}
We neglected the constant $4/\pi$
that is irrelevant to the order estimation.

We numerically examine the effect of nonuniform bias current for pulse separation,
as shown in Fig.~\ref{fig:pulse_separation_total}.
The figure shows numerically obtained snapshots of accrelerated fluxons
by a nonuniform bias current $\gamma (\xi)$,
where its value is $\gamma (\xi) = -0.1$ at the biased region for
$\xi \ge \xi_{\text{b}}$
and $\gamma =0$ for $0 \le \xi < \xi_{\text{b}}$, and
$\xi_{\text{b}} = 15$.
We plot the dynamics of two fluxons with
the same initial velocity $u_0 = 0.01$
and different initial positions of centroid $\xi_0 = 5.75$ and $5.85$,
and thus the spatial width of delay is $\Delta \xi_{\text{d}} = 0.1$.
See Eq.~(\ref{eq:free_kink_expression}) for an explicit expression of the
initial fluxons.
As is seen in the magnified plot at $\tau = 300$ (inset),
the two fluxons are almost completely overlapped before the acceleration.
After the accelaration, the spatial width between the two fluxons are amplified,
and the pluses are contracted.
Whereas the pulses collapse after the acceleration
around $\tau = 410$,
this can be suppressed by increasing the number of discretized points for the numerical calculations,
and thus we expect that the pulse collapse is not observed in an LJJ.

\subsection{\label{subsec:time_delay_detection}Detection of time delay based on SFQ digital circuits}

RSFQ digital circuits operating at several tens of
gigahertz~\cite{likharev1991rsfq} are suitable
for detecting time delays.
The operating frequency of RSFQ circuits roughly
increases in proportion to the square root of the critical current density.
However,
since it becomes difficult to fabricate Josephson junctions with small
critical current values in a high critical current density process,
there is a trade-off with power consumption.
We assume that the appropriate critical current density is
in the range of $0.1$--$1.0\,$kA/cm${}^2$, which allows us to achieve
$5$--$20\,$GHz operation.
We can enhance the time resolution beyond the operation frequency
by exploiting time delay detection techniques,
such as prescaler circuits.
For example, an SFQ time-to-digital converter with
a time resolution of 5 picoseconds or less was demonstrated,
with a critical current density of 1 kA/cm${}^{2}$
at $4.2$ K~\cite{kaplan2001prescaler},
where thermal noise is dominant
at this temperature~\cite{herr2002temperature}.
Hence, we can digitize the information of the qubit state
within low-temperature stages
by detecting the time delay exceeding the time resolution
around 5 picoseconds,
as demonstrated in Fig.~\ref{fig:tr_qubit_time_delay}.
Also, as is estimated by the numerical calculations,
the height of the voltage pulse after the fluxon acceleration is
around one millivolt that is manageable by SFQ digital circuits.
We raise related studies for time delay detection based on SFQ circuits
that we can potentially
employ~\cite{lee1991superconducting,kirichenko2003multi,terai2004timing,nakamiya2007improvement,terabe2007timing,nakamura2009current}.

\section{\label{sec:JTL_multilevel_transmon}Time-delay readout of a multi-level transmon}

\subsection{\label{subsec:multilevel_transmon}Perturbative effect on JTL induced by multi-level transmon}

We now extend the scope to a multi-level transmon
without the two-level approximation.
Recalling the Hamiltonians in Eq.~(\ref{eq:Htr_Hint_quantum_multi})
regarding a multi-level transmon
capacitively coupled with a JTL,
we derive the SW-transformed effective Hamiltonian:
\begin{align}
  \hat{H}_{\text{SW}} &\simeq
  \hat{H}_{\text{tr}}
  - \sum_{n} \frac{ C_{\text{c}}^2 Q_m^2 / (2 C_{\text{J}}^2 C_{\Sigma})}{ [1 - np] [1 - (n+1)p]}
  \ket{n} \! \bra{n},
  \label{eq:H_SW_multi}
\end{align}
where we omit off-diagonal terms that are negligible
for a small-photon-number state $\ket{n}$
(See Appendix~\ref{appsubsec:SW_multi}).
The effective perturbation Hamiltonian commutes with
the free part $\hat{H}_{\text{tr}}$,
and thus the photon-number (Fock) state is not destroyed,
which demonstrates the non-demolition readout~\cite{blais2021circuit}.
The perturbation can be attributed to the change in the capacitance
of the $m$-th junction of the JTL
\begin{align}
  C_{\text{J}} &\mapsto
  C_{\text{J}}
  + \frac{ C_{\text{c}}^2 / C_{\Sigma} }
  {[1 - np] [1 - (n+1) p]},
  \label{eq:C_J_prime_change_multi}
\end{align}
which depends on the photon-number (Fock) state
$\ket{n}$ of the multi-level transmon
without the two-level approximation.
We then derive the dimensionless constant
$\eta_{{\text{c,tr}}} = \eta_{{\text{c,mlt}}} (\ket{n}) $
that depends on the photon number $n$, as
(See Appendix~\ref{appsubsec:perturbation_multi})
\begin{align}
  \eta_{\text{c,mlt}} (\ket{n})
  &= \frac{ -1 }
  {[1 - np] [1 - (n+1) p]}
  \frac{C_{\text{c}}^2 }{ \lambda_{\text{J}} c_{\text{J}} C_{\Sigma}}.
  \label{eq:n_c_tr_ketn_multi}
\end{align}

\subsection{\label{subsec:multilevel_time_delay}Time delay of fluxons}

As is evaluated for a transmon qubit
under the two-level approximation,
we derive an expression for the propagation time $\tau_{\ket{n}}$
compared with that in the absence of coupling $\tau_0$, as
\begin{align}
  \tau_{\ket{n}} - \tau_0
  &\simeq
  \frac{ C_{\text{c}}^2 / (\lambda_{\text{J}} c_{\text{J}} C_{\Sigma}) }
  {[1 - np] [1 - (n+1) p]}
  \frac{1 - u_0^2}{2 u_0}.
\end{align}
We therefore end up with the difference of the propagation
times between the cases for $\ket{n}$ and $\ket{n+1}$, as
\begin{align}
  \tau_{\ket{n}} - \tau_{\ket{n+1}}
  &\simeq
  \frac{- (1 - u_0^2) p C_{\text{c}}^2 / (\lambda_{\text{J}} c_{\text{J}} C_{\Sigma}) }{u_0 [1 - np] [1 - (n+1) p] [1 - (n+2) p]} .
\end{align}
Especially for $n = 0$,
\begin{align}
  T_{\ket{0}} - T_{\ket{1}}
  &\simeq
  - \frac{p}{1 - 2p} \frac{1 - u_0^2}{u_0 (1-p)}
  \frac{C_{\text{c}}^2}{ C_{\Sigma}} Z_{\text{J}},
\end{align}
where the dimension of time is recovored.
Due to the factor $ - p/(1 - 2p)$,
the result does not reproduce the result
in Eq.~(\ref{eq:Td_qubit_ana01})
obtained via the two-level approximation followed by
the SW transformation.
The factor decreases the time delay for a typical transmon as $p \simeq 1/20$,
which is a negative result for one to increase the time delay.
We also face a similar degration of the dispersive shift
due to small anharmonicity of a multi-level transmon~\cite{blais2021circuit}.
The degration of time delay should be compensated by
slowering the fluxon initial vecolity $u_0$
or increasing the characteristic impedance $Z_{\text{J}}$ of the JTL.

\subsection{\label{subsec:multilevel_pinning}Absence of fluxon pinning}

Regarding the fluxon pinning,
the same discussion as that for a transmon qubit
under the two-level approximation can be applied,
and thus the pinning does not occur
even for the multi-level transmon
capacitively coupled to the JTL.

\subsection{\label{subsec:multilevel_nonadiabatic}Nonadiabatic transition}

We evaluate the effects of nonadiabatic transitions
between the transmon energy eigenstates,
namely,
the photon-number (Fock) states $\ket{n}$.
Again, recalling the Hamiltonian of the multi-level transmon
in Eq.~(\ref{eq:Htr_Hint_quantum_multi}),
the transition probability from
$\ket{n_{\text{i}}}$ to $\ket{n_{\text{f}}}$ is calculated as
\begin{align}
  P (n_{\text{f}} | n_{\text{i}})
  &= \frac{1}{\hbar^2}
  \left|
    \int_{-\infty}^{\infty} \! \text{d} s \,
    \nap^{ \frac{\im}{\hbar} (E_{n_{\text{f}}} - E_{n_{\text{i}}}) s }
    \bra{n_{\text{f}}} \hat{V} (s) \ket{n_{\text{i}}}
  \right|^2,
\end{align}
according to Fermi's golden rule.
For the Hamiltonian
$\hat{V} (t) \ceq \im \hbar g_{\text{c}} (t) (\hat{a} - \hat{a}^{\dagger})$,
only transitions between neighbouring photon-number states are allowed,
which is because Fermi's golden rule is second-order perturbative.
The transition probabilities of
$\ket{ n_{\text{i}}} \mapsto \ket{n_{\text{i}} + 1}$
and $\ket{ n_{\text{i}}} \mapsto \ket{n_{\text{i}} - 1}$
are explicitly evaluated, respectively, as
\begin{align}
  P (n_{\text{i}} + 1 | n_{\text{i}})
  &= (n_{\text{i}} + 1)
  \left|
    \int_{-\infty}^{\infty} \! \text{d} s \,
    \nap^{ \im \omega_{n_{\text{i}} + 1} s}
    g_{\text{c}} (s)
  \right|^2, \\
P (n_{\text{i}} - 1 | n_{\text{i}})
  &= n_{\text{i}}
  \left|
    \int_{-\infty}^{\infty} \! \text{d} s \,
    \nap^{ - \im \omega_{n_{\text{i}}} s}
    g_{\text{c}} (s)
  \right|^2,
\end{align}
where $\omega_n = (E_n - E_{n-1}) /\hbar$
denotes the angular frequency of the subspace spanned by
the neighbouring photon-number states $\ket{n}$ and $\ket{n-1}$.
Putting the expression in Eq.~(\ref{eq:explicit_gc_profile})
and exploiting the formula in Eq.~(\ref{eq:sech_Fourier_formula}),
we end up with
\begin{align}
  &P (n_{\text{i}} + 1 | n_{\text{i}}) \n
  &= 2 (n_{\text{i}} + 1)
  \pi^2 n_{0,\text{tr}}^2 \frac{C_{\text{c}}^2}{C_{\Sigma}^2}
  \text{sech}^2
  \left(\frac{\pi}{2} \frac{\sqrt{1 - u_0^2} \omega_{n_{\text{i}} + 1}}
  {u_0 \omega_{\text{p}}}\right) , \\
  &P (n_{\text{i}} - 1 | n_{\text{i}}) \n
  &= 2 n_{\text{i}}
  \pi^2 n_{0,\text{tr}}^2 \frac{C_{\text{c}}^2}{C_{\Sigma}^2}
  \text{sech}^2
  \left(\frac{\pi}{2} \frac{\sqrt{1 - u_0^2} \omega_{n_{\text{i}} }}
  {u_0 \omega_{\text{p}}}\right).
\end{align}
These results resemble the result
in Eq.~(\ref{eq:nonad_transition_prob_qubit}),
while the prefactors $n_{\text{i}} + 1$ and $n_{\text{i}}$
are present due to coefficients of bosonic operators.
As is shown in the discussion of the transmon qubit
under the two-level approximation,
we conclude that the nonadiabatic transitions are negligible
even for a multi-level transmon
when we are interested in the photon-number state $\ket{n}$
that satisfies
\begin{align}
  \frac{\pi}{2} \frac{\sqrt{1 - u_0^2} \omega_{n + 1}}
  {u_0 \omega_{\text{p}}}
  \gg 1.
\end{align}
Thus, decreasing the fluxon initial velocity $u_0$ allows for
the exponential suppression of the nonadiabatic transitions
between the photon-number
states of the multi-level transmon.

\section{\label{sec:radiative}Radiative decay induced by JTL}

\subsection{\label{subsec:KG_transmission_line}Klein--Gordon transmission line (KGTL)}

\begin{figure}
  \centering
		\includegraphics[clip,width=8cm,bb=0 0 710 450]{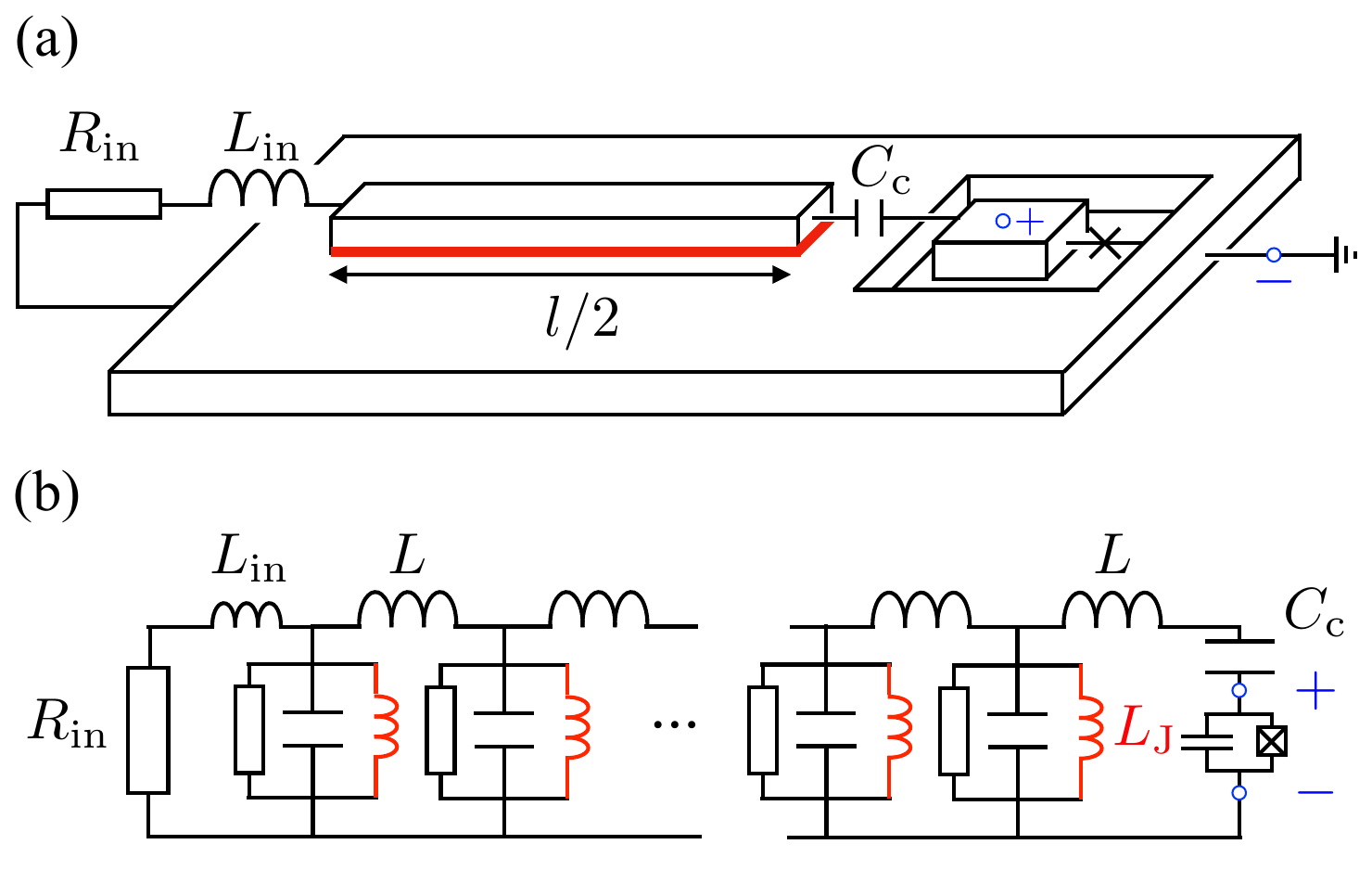}
		\caption{(a) A transmon capacitively coupled to a JTL at one end
    via a coupling capacitor $C_{\text{c}}$,
    and the other end is terminated by an external source made of
    a resistor $R_{\text{in}}$ and an inductor $L_{\text{in}}$.
    The red region represents the JTL
    in which a fluxon propagates.
    (b) A lumped-element circuit diagram of the system.
    The junctions in the JTL are approximated by inductors $L_{\text{J}}$,
    where the Josephson energy $E_{\text{J}}$
    and the corresponding inductance $L_{\text{J}}$
    are associated with each other, as
    $L_{\text{J}} \ceq \varphi_0^2/E_{\text{J}} = \varphi_0 / I_{\text{c}}$
    per unit cell of the JTL.
    The approximated transmission line is called a KGTL.
    The small square resperents a subgap resistor $R_{\text{J}}$
    per unit cell.
    Other parameters are the same as the previous ones.}
		\label{fig:KGTL_LJprime_distributed_lump}
\end{figure}

To evaluate the degree of radiative decay mediated by a JTL,
we work on a transmon capacitively coupled to a JTL at one end,
and the other end is terminated by an external apparatus
controlling input or output signals
(See Fig.~\ref{fig:KGTL_LJprime_distributed_lump}~(a)).
We here rely on a harmonic approximation of the JTL;
we replace the JJs in the JTL by linear inductors,
whose inductance $L_{\text{J}}$ per unit cell
is related to the critical current $I_{\text{c}}$
of the junctions,
as $L_{\text{J}} \ceq \varphi_0 / I_{\text{c}}$
(See Fig.~\ref{fig:KGTL_LJprime_distributed_lump}~(b)).
To take the continuum limit consistently,
we take the limit $\Delta x \to 0$ with the parameter
$\zeta_{\text{J}} \ceq L_{\text{J}} \Delta x$
being constant.
This harmonic approximation is nothing but the truncation of the cosine potential, as
$-\cos \phi \simeq -1 + (1/2) \phi^2$,
and neglecting the constant $-1$.
This is based on the observation that the dynamics of a low-impedance JTL
possessing no fluxon is well captured only by small-amplitude oscillations
around the potential minimum $\phi \simeq0$.
We then obtain the following Klein--Gordon equation to be obeyed:
\begin{align}
  \pd_{\tau}^2 \phi - \pd_{\xi}^2 \phi + \phi
  &= 0,
  \label{eq:dimensionless_KGeq}
\end{align}
where we at the moment omit the dissipation.
Recovering the dimensions of coordinate and time,
the dispersion relation of the Klein--Gordon transmission line (KGTL)
reads
\begin{align}
  \omega^2 (k) &= \omega_{\text{p}}^2 + \overline{c}^2 k^2.
\end{align}
The spectral gap $\omega_{\text{p}}$ is rooted in the mass term $\phi$
appearing in the Klein--Gordon equation in Eq.~(\ref{eq:dimensionless_KGeq}),
and this is reminiscent of a bandgap
in a structured waveguide~\cite{kim2021quantum,jouanny2025high}.
We expect from the dispersion relation that radiative decay
of the qubit having an angular frequency around
$\omega_{\text{q}} \simeq 2 \pi \times 3.7 \,$GHz
is strongly suppressed by the intrinsic spectral gap
$\omega_{\text{p}} \simeq 360 \,$GHz.

\subsection{\label{subsec:admittance_classical_KGTL}Radiative decay induced by KGTL}

We calculate the radiative decay rate induced by a dissipationless KGTL
of length $l/2$ (See
Fig.~\ref{fig:KGTL_LJprime_distributed_lump}~(a)
for an image of the JTL-transmon system and
Fig.~\ref{fig:KGTL_LJprime_distributed_lump}~(b)
for an approximated circuit diagram).
Owing to the electrical circuit theory
(See Appendix~\ref{appsubsec:admittance_radiative}),
we obtain an analytical expression of the radiative decay rate
$\gamma (\omega_{\text{q}})$ of the capacitively coupled transmon at one
end of the KGTL that is terminated by an external impedance
$Z_0 \ceq R_{\text{in}} + \im \omega L_{\text{in}}$
at the other end:
\begin{align}
  \gamma (\omega_{\text{q}})
  &\simeq
  \frac{2 \text{Re}
  \left[
  \im \frac{ \tilde{Z}_{\text{J}} (\omega_{\text{q}}) \{ Z_0^2 + \tilde{Z}_{\text{J}}^2 (\omega_{\text{q}}) \} }
  { \{Z_0 + \im \tilde{Z}_{\text{J}} (\omega_{\text{q}}) \}^2} \right]}
  {C_{\Sigma} \left[\tilde{Z}_{\text{J}} (\omega_{\text{q}}) - 1 /(\omega_{\text{q}} C_{\text{c}})\right]^2}
  \,
  \nap^{-l/\lambda_{\text{J}}}.
  \label{eq:gamma_dissipationless_exp01}
\end{align}
This result demonstrates the exponential suppression of the radiative decay
mediated by the dissipationless KGTL.
We here introduced an effective impedance of KGTL
$\tilde{Z}_{\text{J}} (\omega) \ceq Z_{\text{J}} / \sqrt{ (\omega_{\text{p}}^2 / \omega^2) - 1}$
and assumed the conditions
$\omega_{\text{q}} < \omega_{\text{p}}$
and
$\lambda_{\text{J}} \ll l$
applicable for typical systems.
For a simple expression,
we consider the situation in which
$L_{\text{in}} = 0$ and
$\tilde{Z}_{\text{J}} (\omega_{\text{q}}) \ll R_{\text{in}} , 1/(\omega_{\text{q}} C_{\text{c}})$.
Then, we arrive at
(See Appendix~\ref{appsubsec:dissipationless_KGTL})
\begin{align}
  \gamma (\omega_{\text{q}})
  &\simeq
  4 \omega_{\text{q}}^2 \frac{C_{\text{c}}^2}{C_{\Sigma}}
  \frac{ \tilde{Z}_{\text{J}}^2 (\omega_{\text{q}}) }{ R_{\text{in}} }
  \,
  \nap^{-l / \lambda_{\text{J}}}.
  \label{eq:gamma_rate_dissipationless_KGTL01}
\end{align}

In a practical circumstance,
the JTL contains a finite amount of normal leakage current
described by intrinsic shunt resistors,
also depicted in
Fig.~\ref{fig:KGTL_LJprime_distributed_lump}~(a) and (b).
Under the conditions $\omega_{\text{q}} < \omega_{\text{p}}$
and $\lambda_{\text{J}} \ll l$,
we eventually obtain
\begin{align}
  \gamma (\omega_{\text{q}})
  &\simeq
  \alpha \frac{\omega_{\text{q}}^3}{2 \omega_{\text{p}}}
  \frac{C_{\text{c}}^2}{C_{\Sigma}} \tilde{Z}_{\text{J}} (\omega_{\text{q}}),
  \label{eq:gamma_alpha_main}
\end{align}
which shows the suppression of the radiative decay induced by
an underdamped JTL satisfying $\alpha \ll 1$
(See Appendix~\ref{appsubsec:underdamped}).
For concrete values of parameters in Table~\ref{tab:circuit_params},
we estimate as
$\gamma (\omega_{\text{q}}) \simeq 87\,$mHz
for $\alpha = 1.0 \times 10^{-5}$.
This is negligibly small in typical situations
despite a finite amount of dissipation induced by intrinsic shunt resistors.

\section{\label{sec:discussion}Discussion}

We observed that the time delay is proportional to
the characteristic impedance $Z_{\text{J}}$ of the JTL,
and thus a high-impedance JTL is desirable
for a readout with high accuracy.
One promising approach is the use of high-kinetic-inductance material
for the JTL~\cite{wildermuth2022fluxons},
such as granular Al or NbN thin films~\cite{glezer2020granular,lopez2025superconducting},
to enlarge the quantumness $Z_{\text{J}} / R_{\text{Q}}$
measured by the resistance quantum $R_{\text{Q}} = h / (2e)^2$.
However, as the ratio $Z_{\text{J}} / R_{\text{Q}}$ increases,
quantum fluctuations of the fluxon become significant,
and we cannot ignore its quantum nature any more
and have to treat it as a quantum particle~\cite{kato1996macroscopic,kemp2002josephson,wildermuth2023quantum}.
In such quantum regime of fluxon,
we cannot rely on the classical treatment of the fluxon,
which constitutes the basis of the present paper.
Hence, one intriguing direction to be investigated is
a fully quantum mechanical treatment of the time delay readout,
which we leave for future work.
Beyond the quantum regime,
for $Z_{\text{J}} / R_{\text{Q}} \simeq 1$ or further,
which we might call deep quantum regime of fluxon,
the quantum fluctuations become dominant and
the picture of stable fluxon shall be
violated~\cite{coleman1975quantum,kato1996macroscopic,kemp2002josephson,wildermuth2022fluxons,wustmann2026rapid}.
This is another interesting system to be investigated,
whereas its utility as a readout probe would be limited.

Comparing our result with that for the time-delay readout of a flux qubit
based on the inductive coupling~\cite{fedorov2007reading},
the feature of the absence of fluxon pinning for the capacitive coupling
takes an advantage when slowering the fluxon to obtain a sufficiently large time delay.
On the other hand,
our scaling of the time delay $1/ u_0$ in the limit $u_0 \to 0$
is milder compared with that $1 / u_0^3$ for inductive coupling,
which could be in turn the disadvantage of our scheme.
More quantitatively,
the time delay of the fluxon
induced by a flux qubit inductively coupled to a JTL reads
$\tau_{\text{d}, \text{FQ}} \simeq \phi_{\text{FQ}} / u_0^3$,
where $\phi_{\text{FQ}}$ denotes the dimensionless coupling constant
of the inductive coupling between the flux qubit and the JTL~\cite{fedorov2007reading}.
The pinning velocity $u_{\text{pin}}$
below which the fluxon is pinned is
$u_{\text{pin}} \ceq \sqrt{ \phi_{\text{FQ}} / 2}$~\cite{fedorov2007reading,aslamazov1984pinning}.
In order to enhance the time delay
without suffering from the pinning,
we can not only decrease the initial velocity of the fluxon $u_0$
but also adaptively change the coupling constant $\phi_{\text{FQ}}$,
as $\phi_{\text{FQ}} \propto u_0^2$.
Then, the inequality $u_0 > u_{\text{pin}}$ is always satisfied
even for a slowered velocity $u_0$
and the pinning can be avoided.
Meanwhile, the scaling of the time delay becomes milder, as
$\tau_{\text{FQ}} \propto 1/u_0$.
This result shows that the scaling of the time delay
$\tau_{\text{d}} \propto 1/ u_0$
is common for the capacitive and inductive coupling,
under the condition of the absence of fluxon pinning.

\section{\label{sec:conclusion}Conclusion}

We studied the time-delay readout of a superconducting qubit
coupled with a JTL that serves not only as a
transmission line for a propagating fluxon
but also as a filter
suppressing the radiative decay of the qubit.
We concretely treat a transmon qubit
under the two-level approximation capacitively coupled with a JTL
and obtain analytical and numerical results of the time delay in good agreement,
and the value of the time delay of a sufficiently slow fluxon
in an underdamped JTL exceeds the typical time resolution
of a time-delay detector based on SFQ digital circuits.
Unlike the system of a flux qubit inductively coupled with
a JTL~\cite{fedorov2007reading},
the fluxon pinning is absent for the capacitive coupling,
which is advantageous for the time-delay readout.
Nonadiabatic transitions caused by the fluxon
are suppressed by slowering the fluxon initial velocity,
and thus the use of a sufficiently
slow fluxon enables the non-demolition
and high-fidelity readout.
We also generalize the framework to readout of a multi-level transmon
without the two-level approximation,
while the discrepancy between the analytical results of the time delay
with and without the two-level approximation
is probably due to
virtual excitations to higher excited states.
To evaluate the radiative decay rate of the qubit
induced by the JTL possessing no fluxon,
we approximate JJs in the JTL by linear inductors
to describe small-amplitude oscillations around the minima of cosine potentials,
after which we call the JTL a KGTL.
Then, due to the spectral gap of the KGTL,
we obtain an exponentially suppressed decay rate as a function of distance from
an external environment.
Even for a dissipative JTL having normal current leakage
induced by a finite subgap resistance,
the leading-order contribution to the decay rate is propotional to
the damping coefficient $\alpha$ of the JTL,
and it is also strongly suppressed for an underdamped JTL.
Our readout scheme enables us to circumvent the wiring complexity
and thermal-loaded overheads involved in the brute-force scaling
of superconducting circuits with current technology,
and it therefore opens the way towards a large-scale quantum computer
with heterodeneous quantum-classical architecture
based on the use of the SFQ digital logic family.

\begin{acknowledgments}

We thank Yoshihito Hashimoto and Ryoji Miyazaki for their insightful feedback.
This work was supported by
JST Moonshot R\&D Grant Number JPMJMS2067, Japan.

\end{acknowledgments}

\appendix

\section{\label{appsec:SW_transf}Effective Hamiltonian of transmon weakly coupled to JTL}

\subsection{\label{appsubsec:SW_derivation}Schrieffer--Wolff (SW) transformation}

We begin with the total Hamiltonian
$\hat{H} (t) \ceq \hat{H}_0 + \hat{V} (t)$
composed of an easily diagonalizable free Hamiltonian $\hat{H}_0$
and a small time-dependent perturbation Hamiltonian $\hat{V} (t)$.
For an application of the Schrieffer--Wolff (SW) transformation,
we adopt a time-dependent unitary operator
$\hat{U} (t) \ceq \nap^{- \hat{S} (t)}$
that is determined by an anti-Hermitian operator $\hat{S} (t)$.
Moving onto the rotating frame with respect to $\hat{U}$,
a quantum state is transformed as
$\ket{\psi_U (t)} \ceq \hat{U} \ket{\psi (t)}$
and the transformed Hamiltonian $\hat{H}_U (t)$ reads
\begin{align}
  \hat{H}_U (t) &\ceq
  \im \hbar \frac{\pd \hat{U}}{ \pd t} \hat{U}^{\dagger}
  + \hat{U} \hat{H} \hat{U}^{\dagger}.
\end{align}
In terms of the anti-Hermitian operator $\hat{S} (t)$,
\begin{align}
  \hat{H}_U &=
  - \im \hbar \frac{\pd \hat{S}}{\pd t}
  + \hat{H}_0
  + \hat{V} - [ \hat{S} , \hat{H}_0 ] \n
  & - [\hat{S}, \hat{V}] + \frac{1}{2} \bigl[ \hat{S}, [\hat{S}, \hat{H}_0] \bigr]
  + \frac{1}{2} \bigl[ \hat{S}, [\hat{S}, \hat{V}] \bigr]
  + \cdots,
\end{align}
which is derived from the Baker--Campbell--Hausdorff formula.
Then, we impose the condition on $\hat{S}$ that
\begin{align}
  \hat{V} &= [ \hat{S}, \hat{H}_0],
  \label{eq:SW_transf_cond}
\end{align}
which gives us the unitary transformation called the SW transformation.
Observing that $\hat{S}$ is the same order as
$\hat{V}$ due to the above condition in Eq.~(\ref{eq:SW_transf_cond}),
we arrange the transformed Hamiltonian
$\hat{H}_{\text{SW}} (t) \ceq \hat{H}_U (t)$
with respect to the order of $\hat{V}$
and obtain
\begin{align}
  \hat{H}_{\text{SW}} &=
  - \im \hbar \frac{\pd \hat{S}}{\pd t}
  + \hat{H}_0
  - \frac{1}{2} [ \hat{S}, \hat{V}]
  + O (\hat{V}^3).
\end{align}
Neglecting the higher-order terms $O (\hat{V}^3)$
for a small perturbation Hamiltonian $\hat{V}$,
we obtain the effective Hamiltonian
\begin{align}
  \hat{H}_{\text{SW}} &\simeq
  - \im \hbar \frac{\pd \hat{S}}{\pd t}
  + \hat{H}_0
  - \frac{1}{2} [ \hat{S}, \hat{V}].
\end{align}

In general, we can diagonalize the free Hamiltonian as
$\hat{H}_0 = \sum_n \epsilon_n \ket{\psi_n} \! \bra{\psi_n}$.
Then, the following anti-Hermitian operator $\hat{S} (t)$ satisfies
the condition
of the SW transformation in Eq.~(\ref{eq:SW_transf_cond}):
\begin{align}
  \hat{S} (t) &\ceq
  \sum_{n,m ; \epsilon_n \neq \epsilon_m}
  \frac{ \bra{\psi_m} \hat{V} (t) \ket{\psi_n} }{ \epsilon_n - \epsilon_m}
  \ket{\psi_m} \! \bra{\psi_n}.
  \label{eq:general_hatS_SWtransf}
\end{align}

\subsection{\label{appsubsec:SW_qubit}Two-level approximation}

Under the two-level approximation,
\begin{align}
  \hat{H}_0 &= \frac{\hbar \omega_{\text{q}} }{2} \hat{\sigma}_z, \\
  \hat{V} (t) &= \hbar g_{\text{c}} (t) \hat{\sigma}_y.
\end{align}
The SW transformation $\hat{U}_{\text{SW}} (t) \ceq \nap^{- \hat{S} (t)}$
is given by
$\hat{S} (t) \ceq \im \, (g_{\text{c}} (t) / \omega_{\text{q}}) \hat{\sigma}_x$
that leaves us the transformed Hamiltonian:
\begin{align}
  \hat{H}_{\text{SW}} (t) &=
  \frac{ \hbar \dot{g}_{\text{c}} (t) }{ \omega_{\text{q}} } \hat{\sigma}_x
  + \frac{\hbar \omega_{\text{q}} }{2} \hat{\sigma}_z
  + \frac{\hbar g_{\text{c}}^2 (t) }{\omega_{\text{q}}} \hat{\sigma}_z
  + O (g_{\text{c}}^3).
\end{align}
We obtain the effective Hamiltonian in Eq.~(\ref{eq:SW_Hamiltonian_qubit})
in the main text
by neglecting the higher-order terms $O(g_{\text{c}}^3)$
and the first term in the right-hand side proportional to the time derivative
$\dot{g}_{\text{c}} (t)$,
which is validated for a weak coupling
a small fluxon velocity satisfying
$u_0 \ll \omega_{\text{q}} / \omega_{\text{p}}$.
Although one might worry about the difference of the frame
before and after the SW transformation,
the coupling amplitude $g_{\text{c}} (t)$ is zero
after the fluxon scattering,
and thus we go back to the original frame.

\subsection{\label{appsubsec:SW_multi}Multi-level transmon}

For a multi-level transmon,
the Hamiltonians are
\begin{align}
  \hat{H}_0 &\ceq \sum_{n} E_n \ket{n} \! \bra{n}, \\
  \hat{V} (t) &\ceq - \im \sqrt{2} \mathcal{E} (\hat{a} - \hat{a}^{\dagger}),
\end{align}
where
$\mathcal{E} \ceq e n_{0,\text{tr}} C_{\text{c}} Q_m / (C_{\Sigma} C_{\text{J}}) $.
The SW-transformed Hamiltonian is
$\hat{H}_{\text{SW}}
= \hat{H}_{0} + ( \hat{V} \hat{S} + \text{H.c.})/2$,
where
\begin{align}
  \frac{1}{2 \mathcal{E}^2} \hat{V} \hat{S}
  &= (\hat{a} - \hat{a}^{\dagger})
  \sum_{n,m} \frac{ \bra{m} (\hat{a} - \hat{a}^{\dagger}) \ket{n}}
  {E_m - E_n} \ket{m} \! \bra{n}.
\end{align}
The symbol ``H.c.'' denotes the Hermitian conjugate.
After some calculations, we obtain
$\frac{1}{2} \hat{V} \hat{S} + \text{H.c.} = \hat{D} + \hat{W}$,
where
\begin{align}
  \hat{D} &\ceq
  - \frac{2 \mathcal{E}^2}{\hbar \omega_{\text{tr}}}
  \sum_{n=0}^{\infty} \frac{ 1 } {[1 - np] [1- (n+1) p]} \ket{n} \! \bra{n}, \\
  \hat{W} &\ceq
  - \frac{p \mathcal{E}^2}{ \hbar \omega_{\text{tr}}}
  \sum_{n=0}^{\infty} \frac{ \sqrt{(n+1) (n+2) } }
  {[1 - (n+1) p] [1 - (n+2) p]} \ket{n+2} \! \bra{n} \n
  & + \text{H.c.}
\end{align}
To safely neglect the off-diagonal term $\hat{W}$,
we impose the condition on the photon number $n$ of interest,
that
\begin{align}
  \frac{ 1 - (n+2) p } {p (1 - np) } 
  &\gg
  \frac{1}{2} \sqrt{(n+1) (n+2) }.
\end{align}
For $p = 1/20$, the ratio $Q (n) \ceq L (n) / R (n)$
of the left-hand side
$L (n) \ceq \frac{ 1 - (n+2) p } {p (1 - np) }$
to the right-hand side
$R (n) \ceq \frac{1}{2} \sqrt{(n+1) (n+2) }$
satisfies $Q (n) \ge 10$ for $n =0, 1, 2$,
which validates the effective Hamiltonian
in Eq.~(\ref{eq:H_SW_multi})
up to $n = 2$.
A smaller $p$ validates the effective Hamiltonian for a larger photon number $n$,
while the time delay is propotional to $p$ and thus unfavorably decreases.
Neglecting the off-diagonal term $\hat{W}$,
we obtain $\hat{H}_{\text{SW}} \simeq \hat{H}_0 + \hat{D}$
and recovoer the SW-transformed Hamiltonian
in Eq.~(\ref{eq:H_SW_multi}) in the main text.

\section{\label{appsec:perturbation_sGkink}Perturbation theory of sine-Gordon kink}

\subsection{\label{appsubsec:perturbation_qubit}Perturbation term for a transmon qubit
under the two-level approximation}

We recall the perturbed JTL Hamiltonian
in Eq.~(\ref{eq:H_JTL_prime_twolevel}) in the main text.
The perturbation term can be attributed to the change in the capacitance
of the $m$-th junction of JTL
$C_{\text{J}} \mapsto
C_{\text{J}} + \delta C_{\text{J}} (\sigma_z)$,
where $\delta C_{\text{J}} (\sigma_z)$ depends on the state of qubit as
\begin{align}
  \delta C_{\text{J}} (\sigma_z)
  &\ceq - 4 n_{0,\text{tr}}^2
  \frac{e^2}{\hbar \omega_{\text{q}} }
  \frac{C_{\text{c}}^2}{C_{\Sigma}^2} \sigma_z .
\end{align}
We neglected higher-order terms $O(C_{\text{c}}^4)$.
Remembering the discretized sine-Gordon equation in Eq.~(\ref{eq:sG_discrete_JTL1})
in the main text,
the perturbed fluxon obeys the following equation:
\begin{align}
  & \frac{\varphi_0 C_{\text{J}}}{I_{\text{c}}}
  \frac{\pd^2 \phi_n}{\pd t^2}
	- \frac{\varphi_0}{L I_{\text{c}}} (\phi_{n+1} - 2 \phi_n + \phi_{n-1})
	+ \sin \phi_n \n
  & = -\delta_{n,m} \frac{\varphi_0}{I_{\text{c}}}
  \delta C_{\text{J}} (\sigma_z)
  \frac{\pd^2 \phi_n}{\pd t^2}.
\end{align}
In the continuum limit,
the last term in the right-hand side can be arranged,
as
\begin{align}
  &-\delta_{n,m} \frac{\varphi_0}{I_{\text{c}}}
  \delta C_{\text{J}} (\sigma_z)
  \frac{\pd^2 \phi_n}{\pd t^2} \n
  &= \frac{\delta_{n,m}}{\Delta x} \frac{\varphi_0}{i_{\text{c}}}
  \cdot 4 n_{0,\text{tr}}^2 \frac{e^2}{\hbar \omega_{\text{q}} }
  \frac{C_{\text{c}}^2}{C_{\Sigma}^2} \sigma_z
  \frac{\pd^2 \phi_n}{\pd t^2} \n
  &\to \delta (\xi) \frac{\varphi_0}{\lambda_{\text{J}} i_{\text{c}}}
  \cdot 4 n_{0,\text{tr}}^2
  \frac{e^2}{\hbar \omega_{\text{q}}}
  \frac{C_{\text{c}}^2}{C_{\Sigma}^2} \omega_{\text{p}}^2
  \sigma_z \frac{\pd^2 \phi_n}{\pd \tau^2},
\end{align}
where we used the correspondence
$\delta_{n,m} / \Delta x \to \delta (x) = \lambda_{\text{J}}^{-1} \delta (\xi)$
and $\pd_t = \omega_{\text{p}} \pd_{\tau}$.
Finally, noticing
\begin{align}
  \frac{\varphi_0}{\lambda_{\text{J}} i_{\text{c}}}
  \frac{4 n_{0,\text{tr}}^2 e^2}{\hbar \omega_{\text{q}}}
  \frac{C_{\text{c}}^2}{C_{\Sigma}^2} \omega_{\text{p}}^2
  &= \frac{C_{\text{c}}^2 /( \lambda_{\text{J}} c_{\text{J}} C_{\Sigma}) }{1 - \sqrt{E_{\text{C,tr}} / (8 E_{\text{J,tr}})}},
\end{align}
we reproduce the perturbed sine-Gordon equation in
Eq.~(\ref{eq:perturbed_sG_qubit}) in the main text.

\subsection{\label{appsubsec:perturbation_multi}Perturbation term for a multilevel transmon}

For a multi-level transmon,
the change in the capacitance of the $m$-th junction of
the JTL is given by
Eq.~(\ref{eq:C_J_prime_change_multi}).
We then derive the perturbation term for the sine-Gordon equation,
as
\begin{align}
  &- \delta_{k,m} \frac{\varphi_0}{I_{\text{c}}'}
  \frac{ C_{\text{c}}^2 / C_{\Sigma} }
  {[1 - np] [1 - (n+1) p]}
  \frac{\pd^2 \phi_k}{\pd t^2} \n
  &\to \frac{ - 1 }
  {[1 - np] [1 - (n+1) p]}
  \frac{C_{\text{c}}^2}{\lambda_{\text{J}} c_{\text{J}} C_{\Sigma}}
  \delta (\xi) \pd_{\tau}^2 \phi,
\end{align}
which gives us the dimensionless constant
$\eta_{\text{c,tr}} = \eta_{\text{c,mlt}} (\ket{n})$
in Eq.~(\ref{eq:n_c_tr_ketn_multi}).

\subsection{\label{appsubsec:Deqs_u_Xi}Differential equations for $u(\tau)$ and $\Xi (\tau)$}

According to the perturbation theory of sine-Gordon kinks
provided in Refs.~\cite{keener1977solitons,mclaughlin1978perturbation},
we begin with the following sine-Gordon equation
with a perturbation term $f(\phi, \xi)$:
\begin{align}
  \pd_{\tau}^2 \phi
  - \pd_{\xi}^2 \phi
  + \sin \phi
  &= \gamma - \alpha \pd_{\tau} \phi + f (\phi, \xi).
\end{align}
The perturbation theory is based on the assumptions
that $|\gamma|, \alpha, |f| \ll 1$
under which the soliton shape is preserved
and the perturbed solution of the
dimensionless modulated velocity $u (\tau)$
and the centroid $\Xi (\tau)$ in Eq.~(\ref{eq:modulated_solution_kink})
well approximates the true solution.
Then, for
a perturbation term $f (\phi, \xi)$,
the modulation parameters $u(\tau)$ and $\Xi (\tau)$ obey the following
ordinary differential equations~\cite{keener1977solitons,mclaughlin1978perturbation,fedorov2007reading}:
\begin{align}
  \frac{\text{d} u}{\text{d} \tau} &=
  - \frac{1}{4} \pi \gamma (1 - u^2)^{\frac{3}{2}}
  - \alpha u (1 - u^2) \n
  & \ \ \
  - \frac{1}{4} (1 - u^2)
  \int_{-\infty}^{\infty} \! \text{d} \xi \,
  f (\phi (\xi, \tau), \xi) \, \text{sech} \Theta, 
  \label{eq:du_dtau_ODE_perturbation} \\
  \frac{\text{d} \Xi}{\text{d} \tau} &=
  u -\frac{1}{4} u \sqrt{1 - u^2}
  \int_{-\infty}^{\infty} \! \text{d} \xi \,
  f (\phi (\xi, \tau), \xi) \Theta \, \text{sech} \Theta,
  \label{eq:dXi_dtau_ODE_perturbation}
\end{align}
where we define the function $\Theta = \Theta (\xi, \tau)$ as
\begin{align}
  \Theta (\xi, \tau) &\ceq
  \frac{\xi - \Xi (\tau)}{ \sqrt{1 - u^2 (\tau)}} .
\end{align}

For a JTL capacitively coupled to a transmon,
the perturbation term takes the form
\begin{align}
	f (\phi, \xi) &=
  \eta_{\text{c,tr}} \delta (\xi) \pd_{\tau}^2 \phi (\xi, \tau),
\end{align}
which is applicable both for transmons with and without
the two-level approximation.
The dimensionless constant $\eta_{\text{c,tr}}$ such that
$|\eta_{\text{c,tr}}| \ll 1$ depends on the quantum state of the transmon.
The perturbation term can be expanded as
\begin{align}
  f (\phi, \xi)
  &= \eta_{\text{c,tr}} \delta (\xi)
  \pd_{\tau}^2 \left(
    4 \text{arctan}
  \left[
		\nap^{\Theta}
  \right]
  \right) \\
  &= - \eta_{\text{c,tr}} \delta (\xi)
  \frac{2 u^2 (\tau)}{1 - u^2 (\tau)}
  \text{sech}^2 (\Theta) \sinh (\Theta) \n
  & + O(\eta_{\text{c,tr}}^2),
\end{align}
where we assume the expansion
$u (\tau) = u_0 + O(\eta_{\text{c,tr}})$
around the unperturbed initial velocity $u_0$
and neglect the higher order terms $O (\eta_{\text{c,tr}}^2)$.
Putting this result into
Eqs.~(\ref{eq:du_dtau_ODE_perturbation})
and
(\ref{eq:dXi_dtau_ODE_perturbation})
and executing the integral,
we obtain
\begin{align}
  \frac{\text{d} u}{\text{d} \tau} &=
  - \frac{1}{4} \pi \gamma (1 - u^2)^{\frac{3}{2}}
  - \alpha u (1 - u^2) \n
  & \ \ \
  - \frac{1}{2} \eta_{\text{c,tr}} u^2 \text{sech}^3 (\Theta_0) \sinh (\Theta_0),
  \label{eq:du_dtau_ODE_transmon_qubit} \\
  \frac{\text{d} \Xi}{\text{d} \tau} &=
 u + \frac{1}{2} \eta_{\text{c,tr}}
 \frac{u^3}{\sqrt{1 - u^2}} \Theta_0 \text{sech}^3 (\Theta_0) \sinh (\Theta_0) .
 \label{eq:dXi_dtau_ODE_transmon_qubit}
\end{align}
Here, we introduce the time-dependent function
$\Theta_0 \ceq - \Theta (0, \tau) = \Xi (\tau) / \sqrt{1 - u^2 (\tau)}$,
which also appears in the main text.

\subsection{\label{appsubsec:pert_sol_capa}Analytical solution of $u (\tau)$ for capacitive coupling}
To obtain analytical results,
we assume the following conditions:
\begin{align}
  |\gamma| &\ll |\eta_{\text{c,tr}}| u^2, \\
  \alpha &\ll |\eta_{\text{c,tr}} u|.
\end{align}
In this case, the equation for the modulation parameter
$u(\tau)$ is simplified as
\begin{align}
\frac{\text{d} u}{\text{d} \tau}
  &=
  - \frac{1}{2} \eta_{\text{c,tr}} u^2 \text{sech}^3 (\Theta_0) \sinh (\Theta_0).
  \label{eq:du_dtau_ODE_transmon_qubit_nogamma_alpha}
\end{align}
We set the ansatz
\begin{align}
  \sqrt{1 - u^2} &= \sqrt{1- u_0^2}
  \left[1 + \frac{1}{2} \eta_{\text{c,tr}} F(\Theta_0)\right].
	\label{eq:ansatz_n_ctr}
\end{align}
This is based on the observation that
the velocity is $u_0$ in the absence of transmon (i.e. $\eta_{\text{c,tr}} = 0$),
and we are now interested in the case $|\eta_{\text{c,tr}}| \ll 1$
and
thus the higher-order terms are negligible.
Taking time derivative of the both sides in Eq.~(\ref{eq:ansatz_n_ctr})
and noticing the equation
\begin{align}
	(\pd_{\tau} \Xi) \sqrt{1-u^2}
	+ \frac{ u \Xi }{ \sqrt{1 - u^2} } \pd_{\tau} u
	&= u \sqrt{1 - u^2},
\end{align}
we obtain the following equation for the function $F$:
\begin{align}
	\frac{\text{d} F}{ \text{d} \Theta_0}
	&= \frac{u_0^2}{ \sqrt{1 - u_0^2 }} \,
	\text{sech}^3 (\Theta_0) \sinh (\Theta_0)
	+ O(\eta_{\text{c,tr}}).
\end{align}
We find the analytical solution of the velocity $u(\tau)$,
which satisfies
Eqs.~(\ref{eq:dXi_dtau_ODE_transmon_qubit})
and (\ref{eq:du_dtau_ODE_transmon_qubit_nogamma_alpha})
under the ignorance of higher-order terms $O(\eta_{\text{c,tr}}^2)$:
\begin{align}
  \sqrt{1 - u^2}
  &=
  \sqrt{1 - u_0^2}
  \left[
    1 - \frac{1}{4} \eta_{\text{c,tr}} \frac{u_0^2}{\sqrt{1-u_0^2}} \,
    \text{sech}^2 (\Theta_0)
  \right].
  \label{eq:velocity_perturbation_solution_nc}
\end{align}
Especially for $\eta_{\text{c,tr}} = \eta_{\text{c}} \sigma_z$,
this result recovers Eq.~(\ref{eq:analytical_u_qubit})
under the two-level approximation in the main text.
This result also corresponds to the results in Eqs.~(24) and (33)
in Ref.~\cite{fedorov2007reading}
derived for flux qubits,
while the prefactors and function forms are distinct.
We can also recover analytical expressions
of time delay for multi-level transmon
from this retult, by choosing
$\eta_{\text{c,tr}} = \eta_{\text{c,mlt}} (\ket{n})$
in Eq.~(\ref{eq:n_c_tr_ketn_multi}) in the main text.
We did not assume any condition for the
initial velocity $u_0$,
thus the expression in Eq.~(\ref{eq:velocity_perturbation_solution_nc})
is applicable for an arbitrarily chosen initial velocity $u_0$ of the fluxon,
while only slow fluxons are considered for analytical expressions
in Ref.~\cite{fedorov2007reading}.

\section{\label{appsec:admittance_KGTL}Radiative decay induced by KGTL}

\subsection{\label{appsubsec:admittance_radiative}Admittance and radiative decay rate}

We here derive the admittance of a KGTL
seen by a capacitively coupled transmon at one end,
and the other end is terminated by an external source
(See Fig.~\ref{fig:KGTL_LJprime_distributed_lump}~(a) and (b)).

We set the coordinate $x$ along the KGTL,
whose terminated end is at $x = 0$ and the other
coupled to the transmon is at $x = l/2$,
thus the length of KGTL is $l/2$.
We assume that $Z (x)$ denotes the impedance of the KGTL of length $x$
seen from the end to the ground.
We write down the recurrence relation of the impedance $Z (x + \Delta x)$
for the next KGTL to which we added one unit cell at $x + \Delta x$:
\begin{align}
  Z ( x + \Delta x) &=
  \im \omega L
  + \left[
    \frac{1}{Z (x)} + G_{\text{J}}
    + \frac{1}{\im \omega L_{\text{J}} } + \im \omega C_{\text{J}}
  \right]^{-1}.
\end{align}
In the continuum limit,
we neglect the higher-order terms $O (\Delta x^2)$
and assume
$\{ Z (x + \Delta x) - Z (x) \} / \Delta x \simeq \text{d} Z/ \text{d} x$.
We then obtain the following differential equation of $Z (x)$:
\begin{align}
  \frac{ \text{d} Z}{\text{d} x}
  + X_0 Z^2
  &= \im \omega \ell,
\end{align}
where
\begin{align}
  X_0 &\ceq
  g_{\text{J}} - \im \omega c_{\text{J}}
  \left( \frac{\omega_{\text{p}}^2}{\omega^2} - 1\right).
\end{align}
The definitions of parameters in the continuum limit are presented
in the main text.
This is a type of differential equation called Riccati's
equation, and we can analytically solve it:
\begin{align}
  Z (x) &= \lambda_0
  \frac{ (Z_0 + \lambda_0) \nap^{\lambda_0 X_0 x} + (Z_0 - \lambda_0) \nap^{- \lambda_0 X_0 x}}
  { (Z_0 + \lambda_0) \nap^{\lambda_0 X_0 x} - (Z_0 - \lambda_0) \nap^{- \lambda_0 X_0 x}},
\end{align}
where we introduced a complex parameter
$\lambda_0 \ceq \sqrt{ \im \omega \ell / X_0}$.
Another parameter
$Z_0 \ceq R_{\text{in}} + \im \omega L_{\text{in}}$
denotes the external impedance and gives the boundary condition
$Z (0) = Z_0$.

The transmon qubit is coupled to the KGTL via the caoacitor $C_{\text{c}}$,
and the KGTL is terminated by the external source $Z_0$ at $x = 0$.
Then, effective impedance $Z_{\text{eff}}$ that
the transmon sees reads
\begin{align}
  Z_{\text{eff}} (\omega) &\ceq
  Z_1 (l/2)
  + \im \left[
    Z_2 (l/2) - \frac{1}{\omega C_{\text{c}}}
  \right],
\end{align}
where $Z_1 (x) \ceq \text{Re} [ Z (x)]$ and
$Z_2 (x) \ceq \text{Im} [Z (x)]$.
Evaluating the effective admittance
$Y (\omega) \ceq Z_{\text{eff}}^{-1} (\omega)$,
we can estimate the radiative decay rate $\gamma (\omega_{\text{q}})$
at the transmon angular frequency $\omega = \omega_{\text{q}}$
induced by the externally terminated
KGTL~\cite{esteve1986effect,neeley2008transformed}:
\begin{align}
  \gamma (\omega_{\text{q}}) &\ceq
  \frac{1}{C_{\Sigma}} \text{Re} [Y (\omega_{\text{q}})] \n
  &=
  \left. \frac{1}{C_{\Sigma}}
  \frac{Z_1 (l/2) }{ Z_1^2 (l/2) + [ Z_2 (l/2) - 1 / (\omega C_{\text{c}})]^2 }
  \right|_{\omega = \omega_{\text{q}}} .
  \label{eq:gamma_omegaq_app}
\end{align}

\subsection{\label{appsubsec:dissipationless_KGTL}Decay rate for dissipationless KGTL}

For a dissipationless KGTL,
i.e. $g_{\text{J}} = 0$,
then
\begin{align}
  \lambda_0 X_0 &=
  \frac{1}{\lambda_{\text{J}}}
  \frac{\omega}{\omega_{\text{p}}}
  \sqrt{ \frac{\omega_{\text{p}}^2}{\omega^2} - 1}.
\end{align}
The transmon angular frequency
$\omega = \omega_{\text{q}} \simeq 2 \pi \times 3.7 \,$GHz
and the JTL plasma frequency
$\omega_{\text{p}} \simeq 360 \, $GHz
are typical values, and then
$\omega \ll \omega_{\text{p}}$
and $ \lambda_0 X_0 \simeq 1/\lambda_{\text{J}}$.
As a result, we can approximate the impedance $Z (x)$, as
\begin{align}
  Z (l/2) &=
  \lambda_0
  + 2 \lambda_0 \frac{Z_0^2 - \lambda_0^2}{(Z_0 + \lambda_0)^2}
  \Gamma_0
  + O(\Gamma_0^2),
\end{align}
where $\Gamma_0 \ceq \nap^{ - \lambda_0 X_0 l}$
is an exponentially decaying function with respect to the length $l/2$ of KGTL.
Under the present conditions,
$\lambda_0 = \im \tilde{Z}_{\text{J}}$
is pure imaginary
and thus the real and imaginary parts $Z_1 (x)$ and $Z_2 (x)$
respectively read
\begin{align}
  Z_1 (x) &=
  2 \text{Re} \left[
    \im \tilde{Z}_{\text{J}}
    \frac{Z_0^2 + \tilde{Z}_{\text{J}}^2}{(Z_0 + \im \tilde{Z}_{\text{J}})^2}
  \right]
  \Gamma_0
  + O(\Gamma_0^2), \\
  Z_2 (x) &=
  \tilde{Z}_{\text{J}}
  + O (\Gamma_0),
\end{align}
where
$\tilde{Z}_{\text{J}} (\omega) \ceq
Z_{\text{J}} / \sqrt{ (\omega_{\text{p}}^2 / \omega^2) - 1}$.
The radiative decay rate in Eq.~(\ref{eq:gamma_omegaq_app})
is then written as
\begin{align}
  \gamma (\omega_{\text{q}})
  &= \left. \frac{2 \text{Re} \left[
    \im \frac{\tilde{Z}_{\text{J}} (Z_0^2 + \tilde{Z}_{\text{J}}^2)}
    {(Z_0 + \im \tilde{Z}_{\text{J}})^2}
  \right]}{ C_{\Sigma}
  \left[ \tilde{Z}_{\text{J}}
  - 1/ (\omega C_{\text{c}}) \right]^2}
  \Gamma_0
  \right|_{\omega = \omega_{\text{q}}}
  + O (\Gamma_0^2),
\end{align}
which clearly manifests the exponential suppression of
the radiative decay rate
due to the factor
$\Gamma_0 (\omega_{\text{q}}) \simeq \nap^{- l/\lambda_{\text{J}}}$.
This result yields
Eq.~(\ref{eq:gamma_dissipationless_exp01}) in the main text.
To see an explicit expression,
we assume that $L_{\text{in}} = 0$,
then $Z_0 = R_{\text{in}}$ and we obtain
\begin{align}
  \text{Re}
  \left[
    \im \tilde{Z}_{\text{J}} \frac{Z_0^2 + \tilde{Z}_{\text{J}}^2}{ (Z_0 + \im \tilde{Z}_{\text{J}})^2}
  \right]
  &= \frac{2 R_{\text{in}} \tilde{Z}_{\text{J}}^2 }
  {R_{\text{in}}^2 + \tilde{Z}_{\text{J}}^2} .
\end{align}
This leads us to a simple expression of the radiative decay rate:
\begin{align}
  \gamma (\omega_{\text{q}})
  &\simeq \frac{4}{C_{\Sigma}}
  \frac{1}{\left[\tilde{Z}_{\text{J}} (\omega_{\text{q}}) - 1/(\omega_{\text{q}} C_{\text{c}})\right]^2}
  \frac{ R_{\text{in}} \tilde{Z}_{\text{J}}^2 (\omega_{\text{q}}) }
  {R_{\text{in}}^2 + \tilde{Z}_{\text{J}}^2 (\omega_{\text{q}})}
  \Gamma_0(\omega_{\text{q}}) .
\end{align}
We furthermore consider the situation in which
$1/ (\omega_{\text{q}} C_{\text{c}}) \gg \tilde{Z}_{\text{J}} (\omega_{\text{q}})$,
then
\begin{align}
  \gamma (\omega_{\text{q}})
  &\simeq 4 \omega_{\text{q}}^2 \frac{C_{\text{c}}^2}{C_{\Sigma}}
  \frac{R_{\text{in}} \tilde{Z}^2_{\text{J}} (\omega_{\text{q}}) }{ R_{\text{in}}^2 + \tilde{Z}_{\text{J}}^2 (\omega_{\text{q}})}
  \Gamma_0 (\omega_{\text{q}}),
\end{align}
which reproduces Eq.~(\ref{eq:gamma_rate_dissipationless_KGTL01})
in the main text
under the additional assumption for a low-impedance JTL
$\tilde{Z}_{\text{J}} (\omega_{\text{q}}) \ll R_{\text{in}}$.

\subsection{\label{appsubsec:underdamped}Decay rate for dissipative KGTL}

We here address the radiative decay induced by a dissipative KGTL,
in which a finite amount of dissipation occurs.
In this case, the exponent in the impedance $Z (x)$ reads
\begin{align}
  \lambda_0 X_0 &= \sqrt{ \im \omega \ell X_0} \n
  &= \frac{\omega}{\lambda_{\text{J}} \omega_{\text{p}}}
  \sqrt{ \frac{\omega_{\text{p}}^2}{\omega^2} - 1}
  \left[
    1 + \im \frac{ \alpha \omega_{\text{p}} / \omega }
    { (\omega_{\text{p}}^2/\omega^2) - 1}
  \right]^{1/2} .
\end{align}
We assume that $\alpha \ll 1$ that is validated for an underdamped
(but dissipative) JTL
and $\omega < \omega_{\text{p}}$,
then
$k_1 \ceq \text{Re} [\lambda_0 X_0]$ and
$k_2 \ceq \text{Im} [\lambda_0 X_0]$
are respectively approximated as
\begin{align}
  k_1 &\simeq \frac{\omega}{\lambda_{\text{J}} \omega_{\text{p}}}
  \sqrt{ \frac{\omega_{\text{p}}^2}{\omega^2} - 1}, \\
  k_2 &\simeq \frac{\alpha}{2 \lambda_{\text{J}}}
  \frac{1}{ \sqrt{ (\omega_{\text{p}}^2/\omega^2) - 1}}.
\end{align}
Going back to the expression of $Z (l/2)$,
\begin{align}
  Z (l/2)
  &= \lambda_0 \frac{ (Z_0 + \lambda_0) + (Z_0 - \lambda_0) \nap^{- \im k_2 l} \Gamma_1}
  { (Z_0 + \lambda_0) - (Z_0 - \lambda_0) \nap^{-  \im k_2 l} \Gamma_1},
\end{align}
where $\Gamma_1 \ceq \nap^{-k_1 l}$.
Neglecting the exponentially small factor $\Gamma_1$ for
$k_1 l \simeq l/\lambda_{\text{J}} \gg 1$,
we obtain
\begin{align}
  Z (l/2) &\simeq \lambda_0.
\end{align}
We then estimate the real and imaginary parts of $\lambda_0$, as
\begin{align}
  \text{Re} [\lambda_0] &=
  \frac{\alpha}{2}
  \frac{Z_{\text{J}}}{ \sqrt{(\omega_{\text{p}}^2/\omega^2) - 1} }
  \frac{\omega_{\text{p}} / \omega }{ (\omega_{\text{p}}^2 / \omega^2 ) - 1}
  + O (\alpha^3), \\
  \text{Im} [\lambda_0] &=
  \frac{Z_{\text{J}}}{ \sqrt{(\omega_{\text{p}}^2/\omega^2) - 1} }
  \left[
    1 - \frac{\alpha^2}{2}
    \frac{\omega_{\text{p}}^2 / \omega^2 }{ [(\omega_{\text{p}}^2 / \omega^2 ) - 1]^2}
  \right],
\end{align}
which are justified under the present situation satisfying
$(\alpha \omega_{\text{p}} / \omega) / [(\omega_{\text{p}}^2 / \omega^2) - 1] \ll 1$.
Plugging
$Z_1 = \text{Re}[\lambda_0]$ and $Z_2 = \text{Im} [\lambda_0]$
into Eq.~(\ref{eq:gamma_omegaq_app}),
we arrive at
\begin{align}
  \gamma (\omega_{\text{q}})
  &\simeq \frac{\alpha}{2}
  \frac{\tilde{Z}_{\text{J}} (\omega_{\text{q}}) }
  {C_{\Sigma} [ \tilde{Z}_{\text{J}} (\omega_{\text{q}}) - 1/(\omega_{\text{q}} C_{\text{c}})]^2}
  \frac{ \omega_{\text{p}} / \omega_{\text{q}} }
  { (\omega_{\text{p}}^2 / \omega_{\text{q}}^2) - 1},
\end{align}
where we neglect the higher-order terms $O (\alpha^3)$.
Approximate it further using the relations
$\omega_{\text{q}} C_{\text{c}} \ll \tilde{Z}_{\text{J}}^{-1} (\omega_{\text{q}})$
and
$\omega_{\text{p}} \gg \omega_{\text{q}}$,
we then end up with Eq.~(\ref{eq:gamma_alpha_main}) in the main text.

\nocite{*}

\bibliography{main_SFQ_bib}

\end{document}